%% file: main.tex
\documentclass[sigconf]{acmart}

\usepackage{listings}        
\usepackage[bottom,multiple]{footmisc} 
\usepackage{xspace} 
\usepackage{subcaption} 
\usepackage{caption} 

\newcommand{\nonparam}{nonparametric}
\newcommand{\nonconform}{nonconforming}
\newcommand{\posthoc}{post hoc}
\newcommand{\multifactor}{multifactor}
\newcommand{\typeIerror}{Type I error}
\newcommand{\typeIIerror}{Type II error}

\newcommand{\artcon}{ART-C}

\newcommand{\numdata}{72,000}

\newcommand{\f}[1]{{\textsc{#1}}\xspace}

\usepackage{enumitem}

\setlist[itemize]{leftmargin=*}
\setlist[enumerate]{leftmargin=*,label=\arabic*.}

\AtBeginDocument{%
 \abovedisplayskip=3pt plus 3pt minus 3pt
 \abovedisplayshortskip=3pt plus 3pt minus 3pt
 \belowdisplayskip=3pt plus 3pt minus 3pt
 \belowdisplayshortskip=3pt plus 3pt minus 3pt
}

\AtBeginDocument{%
  \providecommand\BibTeX{{%
    \normalfont B\kern-0.5em{\scshape i\kern-0.25em b}\kern-0.8em\TeX}}}

\acmConference[]{}{}{}
\acmYear{}
\copyrightyear{}
\acmPrice{}
\acmDOI{}
\acmISBN{}
\setcopyright{none}
\begin{document}

\title[ART-C]{An Aligned Rank Transform Procedure for \\Multifactor Contrast Tests}

\author{Lisa A. Elkin}
\affiliation{%
\institution{Paul G. Allen School of Computer Science \& Engineering\\
University of Washington}
\city{Seattle}
\state{WA}
\country{USA}
}
\email{lelkin@cs.washington.edu}

\author{Matthew Kay}
\affiliation{%
\institution{School of Communication\\
Northwestern University}
\city{Evanston}
\state{IL}
\country{USA}
}
\email{mjskay@northwestern.edu}

\author{James. J Higgins}
\affiliation{%
\institution{Department of Statistics\\
Kansas State University}
\city{Manhattan}
\state{KS}
\country{USA}
}
\email{wjhiggins@ksu.edu}

\author{Jacob O. Wobbrock}
\affiliation{%
\institution{
The Information School\\
University of Washington}
\city{Seattle}
\state{WA}
\country{USA}
}
\email{wobbrock@uw.edu}

\renewcommand{\shortauthors}{Elkin, Kay, Higgins, Wobbrock}

\begin{abstract}
\input{content/abstract}
\end{abstract}


\maketitle

\input{content/introduction}

\input{content/related_work}

\input{content/problem}

\input{content/solution}

\input{content/validation}

\input{content/tools}

\input{content/discussion}

\input{content/limitations_and_future_work}

\input{content/conclusion}

\bibliographystyle{ACM-Reference-Format}
\bibliography{references}

\end{document}

%% file: content/abstract.tex
Data from multifactor HCI experiments often violates the normality assumption of parametric tests (i.e., \textit{\nonconform{} data}). \textit{The Aligned Rank Transform (ART)} is a popular nonparametric analysis technique that can find main and interaction effects in nonconforming data, but leads to incorrect results when used to conduct contrast tests. We created a new algorithm called \textit{ART-C} for conducting contrasts within the ART paradigm and validated it on 72,000 data sets. Our results indicate that ART-C does not inflate Type I error rates, unlike contrasts based on ART, and that ART-C has more statistical power than a \textit{t}-test, Mann-Whitney \textit{U} test, Wilcoxon signed-rank test, and ART. We also extended a tool called \textit{ARTool} with our ART-C algorithm for both Windows and R. Our validation had some limitations (e.g., only six distribution types, no mixed factorial designs, no random slopes), and data drawn from Cauchy distributions should not be analyzed with ART-C.

%% file: content/introduction.tex
\section{Introduction}
Statistical procedures are a mainstay of human-computer interaction (HCI) research, particularly in the evaluation of human performance data, like times and error rates; subjective response data, like ordinal ratings and preference indications; and count data, like counts of participants, behaviors, or choices. To improve the quality of results and conclusions drawn from HCI experiments, many in the HCI community have pointed out inadequacies in the methods and tools we use to conduct our statistical analyses and have sought to improve them. For example, Jun et al. \cite{jun_tea_2019} and Wobbrock et al. \cite{wobbrock_aligned_2011} created new tools that make it easier for researchers to conduct analyses correctly, Kay et al. \cite{kay_researcher-centered_2016} and Robertson and Kaptein
 \cite{robertson_modern_2016} introduced the community to modern statistical methods to better serve our needs, and Wobbrock et al. \cite{wobbrock_aligned_2011} and Kaptein et al. \cite{kaptein_powerful_2010} developed statistical methods to better analyze data commonly arising in HCI.
 
Parametric tests such as ANOVA and \textit{t}-test are widely used within HCI, but when experiments give rise to data with residuals that are not normally distributed (i.e., \textit{\nonconform{} data}), researchers and practitioners alike often turn to less familiar \nonparam{} tests. The \textit{Aligned Rank Transform} (\textit{ART}) \cite{higgins_aligned_1990, higgins_1994_aligned,salter_1993_art} is a  \nonparam{} procedure that can detect interaction effects in \multifactor{} experiments. It pre-processes data with an alignment step \cite{hodges_rank_1962} followed by a ranking step \cite{conover_rank_1981}, and the resulting aligned-and-ranked data can be analyzed with an omnibus test, typically an ANOVA. Since its introduction to HCI by Wobbrock et al. \cite{wobbrock_aligned_2011} in 2011, the ART procedure has quickly become a popular technique within HCI, and many HCI venues have published papers that use the ART in their analyses (e.g., \textit{CHI} \cite{amershi_regroup_2012,  gugenheimer_sharevr_2017, hamdan_springlets_2019}, \textit{ASSETS} \cite{azenkot_passchords_2012}, \textit{UIST} \cite{kane_access_2011, roo_one_2017}). Wobbrock et al.'s \textit{ARTool} \cite{wobbrock_aligned_2011} has also been used in publications in several other fields (e.g., cellular biology \cite{ciavardelli_breast_2014}, dentistry \cite{revilla-leon_intraoral_2019}, zoology \cite{feilich_passive_2015}, and cardiology \cite{gaspar_randomized_2018}), and has been cited nearly 900 times thus far.

Although Wobbrock et al. \cite{wobbrock_aligned_2011} mentioned in passing that the original ART's aligning and ranking procedure can be followed by contrast tests, a later R package vignette by Kay \cite{kay_contrast_2020} indicated that contrasts involving combinations of levels across multiple factors (i.e., \textit{\multifactor{} contrasts}) cannot be conducted with the original ART without exploding Type I errors. As it turns out, the data after aligning and ranking for the ART procedure are not properly aligned and ranked for the possible contrast tests that might be conducted. Rather, different alignment-and-ranking procedures must be carried out to enable accurate contrast tests. Our work in this paper contributes an algorithm for proper alignment-and-ranking for contrast tests, and updates open-source ART tools for Windows and R to enable them.

Specifically, we conducted a large-scale analysis to confirm the inappropriateness of using the ART to perform \multifactor{} contrast tests. Inspired by the procedure presented in the aforementioned R package vignette \cite{kay_contrast_2020}, we devised a new procedure for \nonparam{} \multifactor{} contrasts within the ART paradigm: \textit{Aligned Rank Transform Contrasts (\artcon{})}. \artcon{} uses a novel aligning-and-ranking algorithm to pre-process data such that \multifactor{} contrasts can be conducted on the resulting aligned-and-ranked data. To validate ART-C, we created 72,000 synthetic data sets simulating a range of experimental designs, sample sizes, and distribution families and used established statistical simulation procedures \cite{abundis_multiple_2001, blair_comparison_1980, li_robustness_nodate, peterson_six_2002}. We compared the Type I error rate of our new method to a \textit{t}-test, and compared its statistical power to a \textit{t}-test \cite{student_probable_1908}, Wilcoxon signed-rank test \cite{wilcoxon_individual_1945}, Mann-Whitney \textit{U} test \cite{mann_test_1947}, and the original ART \cite{higgins_aligned_1990, higgins_1994_aligned,salter_1993_art}.

Our key findings are that when used to conduct contrasts involving levels from multiple factors, the original ART's Type I error rates are often far from their expected values, and ART's statistical power for is low besides. By comparison, ART-C's Type I error rates are at their expected values and are generally not inflated. Also, for contrasts, ART-C has more statistical power than a \textit{t}-test, Wilcoxon signed-rank test, Mann-Whitney \textit{U} test, and the original ART procedure.

Although ART-C has numerous strengths and general applicability, it should not be used in cases where data appears to have been drawn from a Cauchy distribution (i.e., has Cauchy-distributed residuals). Additionally, the 72,000 data sets created and used in our validation cover a wide range of experimental designs but were not exhaustive. Our synthetic data was limited to two factors with at most three levels each, six types of population distributions, condition sample sizes between 8 and 40, and fully between-subjects or within-subjects designs, not mixed factorial designs. Additionally, ART-C is an alignment-and-ranking procedure to be followed by a contrast test; we chose the \textit{t}-test and did not validate it with other tests.

To facilitate the use of our new \artcon{} procedure, we extended both the open source R version\footnote{R package: https://cran.r-project.org/package=ARTool}\footnote{Code: https://dx.doi.org/10.5281/zenodo.594511} and Windows version\footnote{Windows Tool and Code: http://depts.washington.edu/acelab/proj/art/} of \textit{ARTool}. Both tools are already in widespread use, and our modified versions seamlessly integrate our new \artcon{} procedure for \multifactor{} contrasts. Thus, HCI researchers and others who use either tool can easily use our new versions, and no longer risk incorrectly conducting \multifactor{} contrasts on their aligned-and-ranked data or have to break from the ART paradigm to conduct \multifactor{} contrasts.

Our work contributes: (1) a careful elucidation of the problem of \multifactor{} contrast testing using the ART method described by Wobbrock et al. \cite{wobbrock_aligned_2011}, (2) a new algorithm, \artcon{}, to correctly align-and-rank data for \multifactor{} contrasts within the ART paradigm, (3) validation results from simulation studies showing the correctness and statistical power of our new \artcon{}  procedure, and (4) significant additions to the widely used ARTool R package and ARTool.exe Windows application.

%% file: content/related_work.tex
\section{Related Work}
We created a \multifactor{} contrast testing procedure within the ART paradigm, called ART-C, to enable \multifactor{} contrasts for nonconforming data. Thus, relevant prior research includes the ART procedure itself, the lack of a \multifactor{} contrast testing method within the ART paradigm, and prior statistical contributions directed towards the HCI community.

\subsection{The Aligned Rank Transform}
Rank transforms have been explored in statistics for decades as a basis for nonparametric analyses (e.g. \cite{friedman_use_1937,wilcoxon_individual_1945}). Conover and Iman’s \cite{conover_rank_1981} popular rank transform (RT) procedure applies midranks on responses and then conducts an ANOVA on ranks. However, it was discovered for RT that while Type I error rates for main effects were reasonable, they were unreasonably high for interactions. The aligned rank transform (ART) procedure was developed in response to this problem \cite{higgins_aligned_1990,higgins_1994_aligned,mansouri_multifactor_1998,mansouri_aligned_1999,salter_1993_art}, where responses are first “aligned” \cite{hodges_rank_1962} with respect to the main effect or interaction being analyzed before midranks are applied. The upshot is that both main effects and interactions can be safely analyzed on aligned ranks using ANOVA-type procedures without inflating Type I errors.
Owing to (1) the prevalence of multifactor experiments in HCI, (2) the likelihood of data arising that do not conform to the assumptions of parametric analysis, and (3) the dearth of common statistical procedures to analyze such data, the need for the ART was evident, and in 2011, a paper at CHI was published \cite{wobbrock_aligned_2011} that offered \textit{ARTool}, a Windows application capable of performing data alignment-and-ranking otherwise tedious to conduct by hand. In the decade since, this CHI paper has garnered almost 900 citations according to Google Scholar,\footnote{https://scholar.google.com/scholar?cites=16254127723353600671} indicating the usefulness of ARTool. However, to the best of our knowledge, no prior publication (or tool) has offered a method for conducting contrast tests in the ART paradigm, an essential missing piece particularly after detecting a statistically significant interaction. In this work, we supply this missing piece by devising ART-C and augmenting the open-source ARTool utilities for both Windows and R.

\subsection{Multifactor Contrasts}
Using a single example, Kay \cite{kay_contrast_2020} demonstrated that using the ART to conduct \multifactor{} contrasts can lead to incorrect results; we validate this claim below. A thorough search of the statistics literature did not uncover a suitable solution to the problem of multifactor contrast testing within the ART paradigm. Here we discuss the most closely related statistics work. 

ART contrast methods have been presented in the literature \cite{abundis_multiple_2001,barefield_empirical_2001,mansouri_multifactor_1998,mansouri_aligned_2004}, but the authors did not explain how or whether their methods can be used across multiple factors, showing only examples of single-factor contrasts even in the presence of significant interaction effects. Simulation studies analyzing the effectiveness of ART contrasts also only included data with a single factor \cite{abundis_multiple_2001, barefield_empirical_2001, peterson_six_2002}.

Mansouri et al. \cite{mansouri_multifactor_1998} developed ART analogues of well-known contrast procedures (Tukey's HSD, Scheff\'e's method, Fisher's least significant difference procedure) for data with two factors. However, they did not specify or demonstrate whether their methods are applicable to \multifactor{} contrasts; nonetheless, devising ART analogues to complex contrast procedures is not our objective here. Rather, we sought to find an aligning-and-ranking procedure that can be followed by a common contrast test, especially the familiar \textit{t}-test.

Peterson et al. \cite{peterson_six_2002} compared the effectiveness of the ART using six different statistics in the alignment process (sample mean, sample median, lightly trimmed Winsorized mean, heavily trimmed Winsorized mean, Huber M-estimator, and Harrel-Davis estimator of the median). Rather than changing the type of statistic used for alignment, our method changes the alignment process itself, distinguishing it from Peterson et al.'s work. The authors also did not test whether their methods can be used to analyze \multifactor{} contrasts.

\subsection{Statistics Work in HCI}
Statistical analyses are powerful, but only meaningful when statistical tests are used correctly. HCI researchers are well positioned to improve the quality of results drawn from statistical analyses by looking at them through a usability lens. Tools that aid researchers in using statistical tests correctly can improve the quality of our quantitative practices. Wobbrock et al. \cite{wobbrock_aligned_2011} argued for the importance of \nonparam{} tests that are as easy to use as ANOVA when they extended the ART procedure to multiple factors and provided a tool for carrying it out. Their \textit{ARTool} Windows application and ARTool R package made the ART easy to use, in HCI and beyond.

Other researchers in HCI have also recognized the value of providing useful tools for statistical analysis. Jun et al. \cite{jun_tea_2019} provided \textit{Tea}, a system in which users specify their study design and hypotheses at a high level, and then Tea figures out which tests to run, runs them, and returns the results, lowering the barrier to performing valid statistical tests. Kay et al. \cite{kay_researcher-centered_2016} took a different approach to user-centered statistics and looked at how using Bayesian analysis can help the HCI community accrue knowledge without having to conduct inconvenient larger studies or replication studies, which conflict with the priority the community places on novelty. 

Another important aspect of usable statistics is their visibility and framing. Kay et al. \cite{kay_researcher-centered_2016}, Wobbrock et al. \cite{wobbrock_aligned_2011}, and Kaptein et al. \cite{kaptein_powerful_2010} introduced methods from other fields into the HCI literature. Robertson and Kaptein's \cite{robertson_modern_2016} book, \textit{Modern Statistical Methods for HCI}, introduced the HCI community to current statistical methods. None of these methods were wholly new, but framing them in an HCI context and curating them into an HCI book made them accessible to an HCI audience, which might not otherwise discover them.

Our work introduces the HCI community to our new method, \artcon{}, and an updated version of ARTool for both Windows and R that make \multifactor{} contrasts on \nonconform{} data easier to conduct, lowering the barrier to performing correct statistical analyses within HCI and beyond.

%% file: content/problem.tex
\section{THE PROBLEM: MULTIFACTOR CONTRASTS IN ART}
\begin{figure}[!h]
\centering
\includegraphics[width=\columnwidth]{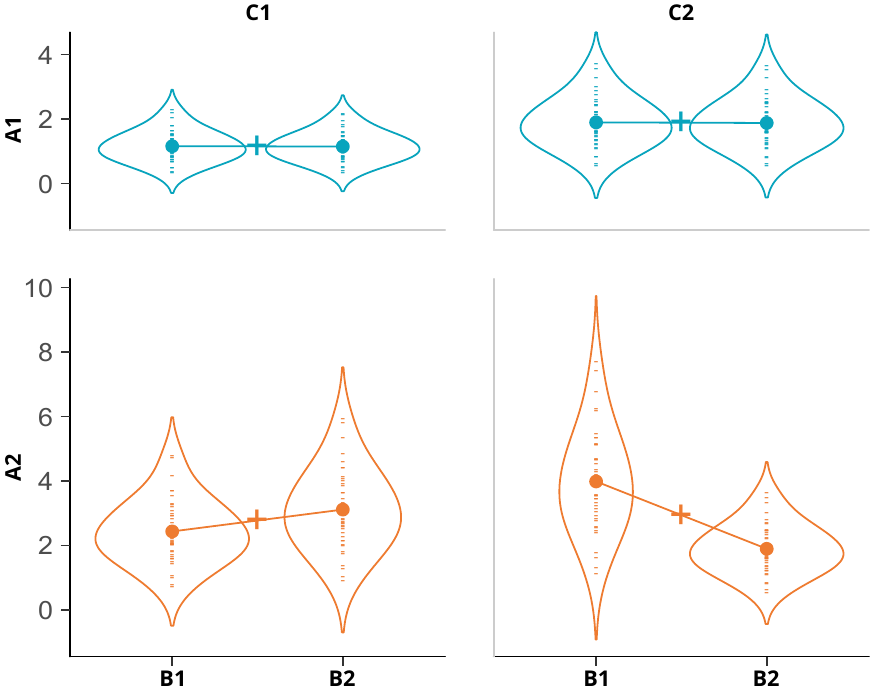}
\caption{Sample data for each condition in our running example. Dots indicate condition means, lines connect condition means for visual comparison, and a plus indicates the mean of both connected conditions.}
\label{fig:eg_all}
\end{figure}

The problem we address in this work is best explained with an example. We refer to this example as our \textit{running example} throughout this paper. Let's consider a within-subjects experiment with three factors having two levels each: $A: \{A1, A2\}$, $B: \{B1, B2\}$, $C: \{C1, C2\}$, and response $Y$. There are 40 subjects and data for each condition is drawn from a log-normal distribution. Table \ref{tab:eg_true_pop_means} shows the log-scale true population means for each condition, and Figure \ref{fig:eg_all} shows the resulting sample data.

\begin{table}[h]
    \caption{Log-scale population means for each condition in our running example.}
    \input{tables/Running_Eg/Eg_True_Pop_Means}
    \label{tab:eg_true_pop_means}
\end{table}

Suppose we analyze this data using the ART. A significant main effect of $A$ would tell us that the level of $A$, i.e., $A1$ \textit{vs}. $A2$, has an effect on the value of $Y$. A significant $A \times B$ interaction would tell us that the effect $A$ has on the value of $Y$ is statistically significantly different for different levels of $B$. Indeed, the original ART procedure works very well for detecting main effects and interactions. But it lacks a suitable method for contrast tests. Contrast tests can tell us which levels of each factor cause these effects; they are commonly used to conduct \textit{post hoc} pairwise comparisons following a statistically significant main effect or interaction, but they can also be used to compare levels of factors directly when warranted by the research question (i.e., planned contrasts). We use the term \textit{single-factor contrasts} to refer to comparisons between levels within a single factor (e.g., \textit{post hoc} tests following a main effect), and \textit{multifactor contrasts} to refer to comparisons between combinations of levels from multiple factors (e.g., \textit{post hoc} tests following a significant interaction effect).

Single-factor contrasts can be conducted safely using a \textit{t}-test on data that has been aligned-and-ranked with the original ART procedure. However, conducting \multifactor{} contrasts on data that has been aligned-and-ranked with the original ART procedure produces incorrect results. 

We show this using our running example. Since we know the data is drawn from a lognormal distribution, we fit a \textit{linear mixed model} (\textit{LMM}) \cite{frederick_fixed-_1999,
ware_linear_1985} to log-transformed data as a baseline, and fit an ART model to the original (not log-transformed) data. Specifically, we wish to compare levels in $A$ and $B$, averaging over the levels of $C$. That is, $C$ is not directly involved in the contrasts. This is achieved in R using the following code.

\begin{lstlisting}
# Fit linear mixed model (LMM)
# to log-transformed data
m.lmm = lmer(log(Y) ~ A*B*C + (1|S), 
    data=df)

# Fit ART model to data
m.art = art(Y ~ A*B*C + (1|S), df)

# Conduct A x B contrasts on LMM
contrast(emmeans(m.lmm, ~ A:B),
    method="pairwise", adjust="holm")

# Conduct A x B contrasts on ART model
contrast(emmeans(
    artlm(m.art, "A:B"), ~ A:B),
    method="pairwise", adjust="holm")
\end{lstlisting}

\begin{figure}[!h]
\centering
\includegraphics[width=\columnwidth]{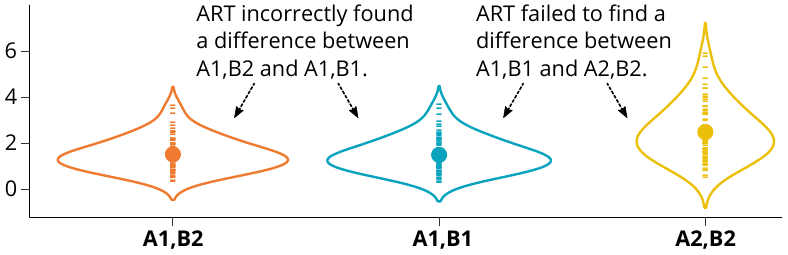}
\caption{Sample data for ($A1,B2$), ($A1,B1$), and ($A2,B2$). Dots indicate condition means. Contrasts using the ART procedure found a difference between $A1,B2$ ad $A1,B1$ even though there is not a true difference; also, no difference was found between $A1,B1$ and $A2,B2$ even though there \textit{is} a true difference.}
\label{fig:eg_all_results}
\end{figure}

We created the data such that there is not a true difference between ($A1,B1$) and ($A1,B2$), and there \textit{is} a true difference between ($A1,B1$) and ($A2,B2$) (Table \ref{tab:eg_true_pop_means}). Comparing results of both tests, contrasts on the LMM produce results that match the true effects ($A1,B1 - A1,B2: \ p = .1792$, i.e., no significant difference) and ($A1,B1 - A2,B2: \ p < .0001$, i.e., a significant difference) (Table \ref{tab:lmm_contrasts}), but contrasts conducted on the ART model result in a \textit{\typeIerror{}} (i.e., finding a significant difference when there is no true difference) $(A1,B1 - B1,B2: \ p < .0001)$, and a \textit{\typeIIerror{}} (i.e., not finding a significant difference when there is a true difference) $(A1,B1 - A2,B2: \ p = .9144)$ (Table \ref{tab:ART_contrasts}, Figure \ref{fig:eg_all_results}).
Both tests' results agree for all other pairs of conditions.

\begin{table}[h]
    \caption{Highlighted results of contrasts conducted on a LMM of log-transformed responses, comparing levels of $A$ and $B$ in our running example. In the top row, a difference was correctly not detected between $A1,B1$ and $A1,B2$ ($p = .1792$), and there is no true difference. In the bottom row, a difference was correctly detected between $A1,B1$ and $A2,B2$ ($p < .0001$), and there \textit{is} a true difference.}
    \input{tables/Running_Eg/lmm_contrasts.tex}
    \label{tab:lmm_contrasts}
\end{table}

\begin{table}[h]
    \caption{Highlighted results of contrasts conducted using ART, comparing levels of $A$ and $B$ in our running example. In the top row, a difference was incorrectly detected between $A1,B1$ and $A1,B2$ ($p < 0.0001$), but there is no true difference. In the bottom row, a difference was incorrectly \textit{not} detected between $A1,B1$ and $A2,B2$ ($p = .9144$), but there \textit{is} a true difference.}
    \input{tables/Running_Eg/ART_contrasts.tex}
    \label{tab:ART_contrasts}
\end{table}

\begin{figure}[!h]
\centering
\Description[ART-C Type I error vs ART Type I error using histograms]{Stacked histograms comparing ART Type I error rates to ART-C Type I error rates for each population distribution. ART-C Type I error rates are are closely clustered around .05 and ART Type I error rates are not as close, particularly for lognormal and exponential distributions.}
\includegraphics[width=\columnwidth]{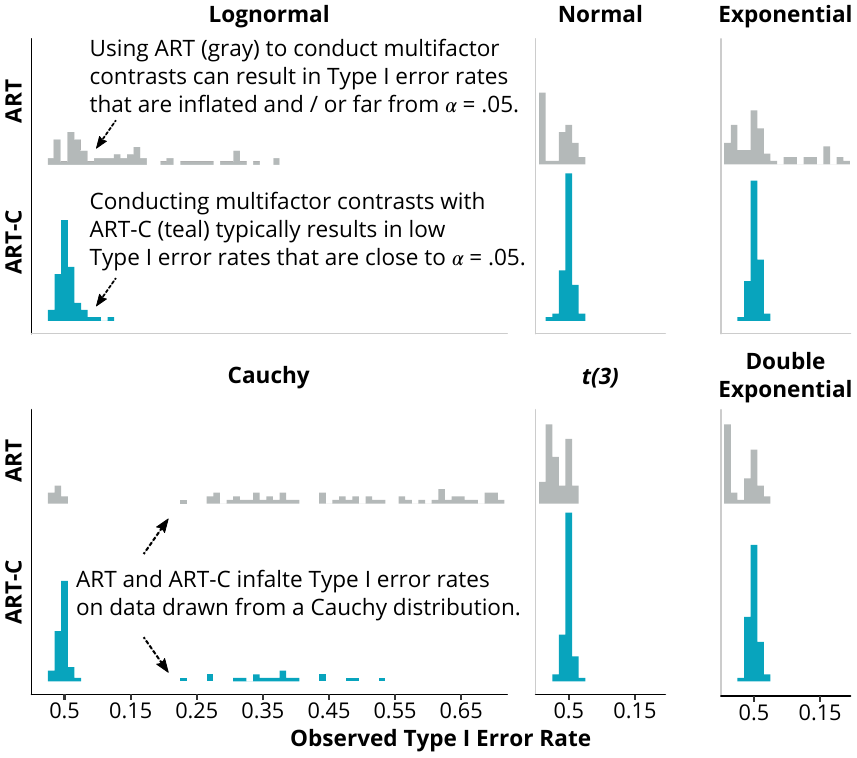}
\caption{ART Type I error rates (gray) compared to ART-C Type I error rates (teal). Each data point represents the observed Type I error rate of one "design," explained below. Values closer to $\alpha=.05$ are better, indicating greater correctness. ART-C Type I error rates are closer to .05 for all distributions.}
\label{fig:ART_TypeI_Distr}
\end{figure}

\begin{figure}[!h]
\centering
\includegraphics[width=\columnwidth]{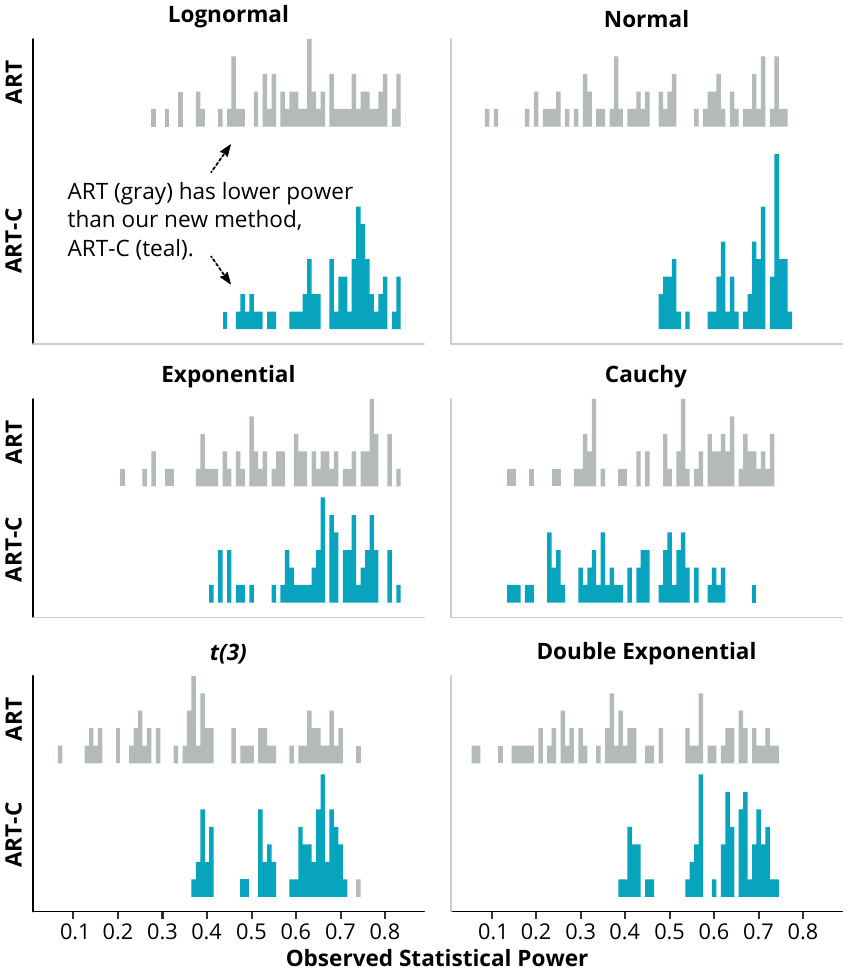}
\Description[ART-C power vs. ART power using histograms]{Stacked histograms comparing ART power to ART-C power for each population distribution.}
\caption{ART statistical power (gray) compared to ART-C statistical power (teal). Each data point represents the observed statistical power of one "design," explained below. Larger values are better, indicating greater power. ART-C has greater power for all distributions except Cauchy.}
\label{fig:ART_Power_Distr}
\end{figure}

Obviously, we cannot judge the validity of a statistical procedure on one example alone. Therefore, we assessed the correctness of the original ART procedure applied to multifactor contrasts on 72,000 synthetic data sets representing several different experimental designs, and confirmed the patterns found in our running example. (More details on our simulation procedure are given below.) Our results show that using the the original ART procedure to conduct multifactor contrasts on data drawn from lognormal, Cauchy, or exponential distributions produces inflated Type I error rates (Figure \ref{fig:ART_TypeI_Distr}); that conducting contrasts with the the original ART procedure on data drawn from \textit{any} distribution produces Type I error rates that are far from their expected value ($\alpha = .05$), either too high or too low (Figure \ref{fig:ART_TypeI_Distr}); and that, for multifactor contrasts, the original ART procedure has low statistical power (high Type II error rates) (Figure \ref{fig:ART_Power_Distr}). Single-factor contrasts conducted with the the original ART procedure are, in fact, correct; indeed, our new method \artcon{} reduces to the same mathematical formula as the original ART in the single-factor case. \artcon{} results are included in Figures \ref{fig:ART_TypeI_Distr} and \ref{fig:ART_Power_Distr} for comparison. We show \artcon{}'s derivation and validation below.

%% file: tables/Running_Eg/Eg_True_Pop_Means.tex
\begin{center}
\begin{tabular}{l|l|l|l|l}
\toprule
\multicolumn{1}{c}{\textbf{Condition}} &
\multicolumn{1}{|c}{\textbf{A}} &
\multicolumn{1}{|c}{\textbf{B}} &
\multicolumn{1}{|c}{\textbf{C}} &
\multicolumn{1}{|c}{\textbf{\begin{tabular}[c]{@{}c@{}}Log-scale\\Population Mean\end{tabular}}}\\
\midrule
1 & A1         & B1         & C1         & 0.00                       \\
\midrule
2 & A1         & B1         & C2         & 0.50                     \\
\midrule
3 & A1         & B2         & C1         & 0.00                       \\
\midrule
4 & A1         & B2         & C2         & 0.50                     \\
\midrule
5 & A2         & B1         & C1         & 0.75                    \\
\midrule
6 & A2         & B1         & C2         & 1.25                    \\
\midrule
7 & A2         & B2         & C1         & 1.00                     \\
\midrule
8 & A2         & B2         & C2         & 0.50  
\\
\bottomrule
\end{tabular}
\end{center}

%% file: tables/Running_Eg/lmm_contrasts.tex
\begin{center}
\begin{tabular}{lrrrrl}
  \hline
contrast & estimate & SE & df & t.ratio & p.value \\ 
  \hline
  A1 B1 - A1 B2 & 0.0 & 0.0 & 273 & 1.3 & 0.1792 \\
  A1 B1 - A2 B2 & -0.5 & 0.0 & 273 & -165.9 & $<$.0001 \\ 
   \hline
\end{tabular}
\end{center}

%% file: tables/Running_Eg/ART_contrasts.tex
\begin{center}
\begin{tabular}{lrrrrl}
  \hline
contrast & estimate & SE & df & t.ratio & p.value \\ 
  \hline
  A1 B1 - A1 B2 & -43.2 & 3.8 & 273 & -11.3 & $<$.0001 \\ 
  A1 B1 - A2 B2 & 0.4 & 3.8 & 273 & 0.1 & 0.9144 \\
   \hline
\end{tabular}
\end{center}

%% file: content/solution.tex
\section{THE SOLUTION: AN ALGORITHM FOR ART CONTRASTS}

We developed \textit{\artcon{}}, a procedure to conduct \nonparam{} \multifactor{} contrasts within the ART paradigm. Using our new alignment process, ART-C aligns data, then ranks it with ascending midranks, and conducts \multifactor{} contrasts comparing combinations of levels of factors for which the data was aligned. Thus, the process is much like the original ART procedure, but the data is aligned not for main effects and interactions, but for desired contrast tests.

Data must be aligned and ranked for each set of factors whose levels will be compared. In our running example, we found an $A \times B$ interaction effect. Response $Y$ must be aligned and ranked to compare combinations of levels of $A$ and $B$. Had we also found an $A \times C$ interaction effect, response $Y$ would have to be aligned and ranked separately to compare combinations of levels of $A$ and $C$.

Like the original ART procedure, ART-C can be used on \nonconform{} data, i.e., data for which the residuals do not need to be normally distributed. (Although, do note that the original ART procedure has been shown to inflate Type I error rates on heteroscedastic data \cite{richter_nearly_1999}.) In this work, we validated ART-C on data with continuous responses.

\subsection{ART-C Procedure for Multifactor Contrasts}

In this section, we walk through the \artcon{} procedure with an example, similar to our running example, with three factors: $A$ with levels $A_i, i=1...a$, $B$ with levels $B_j, j=1...b$, and $C$ with levels $C_k, k=1...c$, and response $Y$.
\\
We present the \artcon{} procedure in four steps:

\smallskip
\noindent \textbf{Step 1. Prepare data.} To prepare data for \artcon{}:
\begin{enumerate}
\item  Concatenate the factors of interest to create a new factor. For example, when conducting contrasts on $A$ and $B$, we concatenate $A$ and $B$ and create a new factor labeled \textit{AB}. For any response $Y$ for which $A$ has level $A_i$ and $B$ has level $B_j$, $AB$ has level $AB_{ij}$.

\item Remove original copies of the factors involved (here $A$ and $B$).

\item Keep as-is any factors not involved in the desired contrasts (here $C$).
\end{enumerate}

\smallskip
\noindent \textbf{Step 2. Compute aligned response $Y'$.} Regardless of which factors were concatenated in Step 1.1 and which original factors were removed in Step 1.2, $Y_{ijk}$ denotes all responses $Y$ where $A$ had level $A_i$, $B$ had level $B_j$, and $C$ had level $C_k$ \textit{before} Step 1.1 was completed. Sometimes, the levels of all factors need to be taken into account when aligning $Y_{ijk}$. $\overline{AB_{ij}C_{k}}$ denotes the mean of all responses where the new concatenated factor $AB$ has level $AB_{ij}$ and factor $C$ has level $C_k$. 
Other times, we only care about the levels of the concatenated factor. $\overline{AB_{ij}}$ denotes the mean of all responses where $AB$ has level $AB_{ij}$ regardless of the level of $C$.  $\mu$ denotes the grand mean (i.e., the mean of all responses).

In our running example, there are three possible types of contrasts: three-factor contrasts, two-factor contrasts, and single-factor contrasts. We demonstrate the ART-C aligning formula on all three types of contrasts and then present the general case. As an example, Table \ref{tab:simple_sample_align} shows a small subset of sample calculations for two-factor contrasts in a three-factor model.

\noindent \textbf{Three-factor contrasts in a three-factor model.}
To align response $Y_{ijk}$ for contrasts between levels of factors $A$, $B$, and $C$, compute:
\begin{align*}
Y_{ijk}' &= Y_{ijk} - \overline{ABC_{ijk}} + \overline{ABC_{ijk}} - \mu\\
&= Y_{ijk} - \mu.
\end{align*}
\noindent \textbf{Two-factor contrasts in a three-factor model.}
To align response $Y_{ijk}$ for contrasts between levels of factors $A$ and $B$, compute:
\begin{align*}
Y_{ijk}' = Y_{ijk} - \overline{AB_{ij}C_{k}} + \overline{AB_{ij}} - \mu
\end{align*}
\noindent \textbf{Single-factor contrasts in a three-factor model.} The focus of this work is not on single-factor contrasts since the original ART alignment procedure can be used for single-factor contrasts, but it is worth noting that our method is mathematically equivalent to the ART in the single-factor case. To align response $Y_{ijk}$ for contrasts between levels of factor $A$ compute:
\begin{align*}
Y_{ijk}' = Y_{ijk} - \overline{A_{i}B_{j}C_{k}} + \overline{A_i} - \mu
\end{align*}
\noindent \textbf{General Case: $M$-factor contrasts in an $N$-factor model.}
We need more complex notation to describe the general case. In the general case, we align response $Y_{ij\dots n}$ for contrasts between levels of $M$ factors in an $N$-factor model. In the example above, we named our factors $A$, $B$, and $C$. Here, we name them $X_1$, $X_2$, ..., $X_N$ and denote level $j$ of factor $X_i$ as $X_{i,j}$ (e.g., level 2 of factor $X_1$ is denoted as $X_{1,2}$). In Step 1.1, we concatenated the $M$ factors for which we were aligning to create a new factor $X_1X_2\dots X_M$. The level of factor $X_1X_2\dots X_M$ that was created by concatenating $X_{1,i}$, $X_{2,j}$, ..., $X_{M,m}$ is denoted as $(X_1X_2\dots X_M)_{ij\dots m}$. In Step 1.2, we removed the original copies of the $M$ factors concatenated in Step 1.1. After Step 1.3, there are $N-M$ other (not concatenated) factors in the model denoted $X_{M+1}, X_{M+2}, \dots , X_N$. Thus, $X_{M+1,m+1}$ denotes a level of factor $X_{M+1}$;  $X_{M+2,m+2}$ denotes a level of factor $X_{M+2}$; and $X_{N,n}$ denotes a level of factor $X_N$. With this notation in hand, to align the data in the general case, we compute:
\begin{multline*}
Y_{ij\dots n}' = Y_{ij\dots n} - \overline{(X_1X_2\dots X_M)_{ij\dots m}
X_{M+1,m+1}X_{M+2,m+2}\dots X_{N,n}}\\
+ \overline{(X_1X_2\dots X_M)_{ij\dots m}} - \mu
\end{multline*}

For example, with this notation, our "two-factor contrasts in a three-factor model" formula would be:
\begin{align*}
Y_{ijk}' = Y_{ijk} - \overline{(X_1X_2)_{ij}X_{3,k}}
+ \overline{(X_1X_2)_{ij}} - \mu
\end{align*} 

\smallskip
\noindent \textbf{Step 3. Compute ranked response $Y''$.}
Apply midranks to all values $Y'$ in ascending order to create aligned-and-ranked responses $Y''$ (see example in Table \ref{tab:simple_sample_align}). That is, the smallest $Y'$ is given rank $Y''$ = 1, the next smallest $Y'$ is given rank $Y''$ = 2, until all $Y'$ values have been assigned a rank. If there is a tie among $k$ values, the mean of the next $k$ ranks that would have been assigned is used as the rank for all $k$ values (i.e., midranks). For example, if there is a tie between the third and fourth smallest $Y'$, they would both be assigned rank $Y''$ = (3 + 4) / 2 = 3.5. This is a standard application of applying ascending midranks to data.
\begin{table}[h]
    \caption{Sample calculations to compute aligned response Y$'$ and aligned and ranked response Y$''$ for two-factor contrasts in a three-factor model using the ART-C procedure. Only 4 of 8 conditions are shown here for considerations of space.}
    \input{tables/simple_sample_align}
    \label{tab:simple_sample_align}
\end{table}

\smallskip
\noindent \textbf{Step 4. Conduct contrasts on $Y''$.}
Contrast tests as \textit{\posthoc{}} tests are justified when the original ART procedure resulted in significant main effects or interactions. However, as stated above, contrasts do not need to follow significant omnibus tests if warranted by the research question (i.e., planned contrasts). Also, note that conducting multiple \textit{post hoc} tests should be accompanied by a \textit{p}-value correction for multiple comparisons (e.g., with the Bonferroni correction \cite{weisstein2004bonferroni}, Holm's sequential Bonferroni procedure \cite{holm1979simple}, or Tukey's HSD test \cite{tukey1949comparing}, to name a few). In the ART-C procedure, conduct contrasts using the full factorial model comprising all factors that remain after Step 1.3, but only interpret results of comparisons between levels of the concatenated factor created in Step 1.1; comparisons between levels of the non-concatenated factor(s) are meaningless.

Returning to our running example of conducting contrasts to compare levels of $A$ and $B$, we have factors $AB$ and $C$, and have computed $Y''$ aligned-and-ranked for factor $AB$. We would therefore conduct contrasts using a full-factorial model with factors $AB$ and $C$ (e.g., $Y \sim AB*C$). We ignore the omnibus test results for this model, but we follow it with contrasts among desired levels of $AB$. Contrasts that would involve $C$ are meaningless.

%% file: tables/simple_sample_align.tex
\begin{center}
\begin{tabular}{l|l|l|l|l}
  \toprule
  \multicolumn{1}{c}{\textbf{AB}} &
  \multicolumn{1}{|c}{\textbf{C}} &
  \multicolumn{1}{|c}{\textbf{Y}} &
  \multicolumn{1}{|c}{\textbf{Y$'$}} &
  \multicolumn{1}{|c}{\textbf{Y$''$}}\\
  \midrule
$AB_{11}$ & $C_1$ & 7 & 7 - $\frac{7+5}{2}$ + $\frac{7+5+2+2}{4}$ - 5 = 0 & 5.5\\
  \midrule
$AB_{11}$ & $C_1$ & 5 & 5 - $\frac{7+5}{2}$ + $\frac{7+5+2+2}{4}$ - 5 = -2 & 1\\ 
  \midrule
$AB_{11}$ & $C_2$ & 2 & 2 - $\frac{2+2}{2}$ + $\frac{7+5+2+2}{4}$ - 5 = -1 & 3\\ 
  \midrule
$AB_{11}$ & $C_2$ & 2 & 2 - $\frac{2+2}{2}$ + $\frac{7+5+2+2}{4}$ - 5 = -1 & 3\\ 
  \midrule
$AB_{12}$ & $C_1$ & 10 & 10 - $\frac{10+8}{2}$ + $\frac{10+8+5+1}{4}$ - 5 = 2 & 7\\ 
  \midrule
$AB_{12}$ & $C_1$ & 8 & 8 -  $\frac{10+8}{2}$ + $\frac{10+8+5+1}{4}$ - 5 = 0 & 5.5\\
  \midrule
$AB_{12}$ & $C_2$ & 5 & 5 - $\frac{5+1}{2}$ + $\frac{10+8+5+1}{4}$ - 5 = 3 & 8\\ 
  \midrule
$AB_{12}$ & $C_2$ & 1 & 1 - $\frac{5+1}{2}$ + $\frac{10+8+5+1}{4}$ - 5 = -1 & 3\\ 
   \bottomrule
\end{tabular}
\end{center}

%% file: content/validation.tex
\section{VALIDATING OUR APPROACH}
In this section, we describe how we validated our ART-C procedure for multifactor contrasts. Specifically, we examined Type I error rates and statistical power. We approached our validation in ways consistent with simulation-based validations from the statistics literature \cite{abundis_multiple_2001, blair_comparison_1980, li_robustness_nodate, peterson_six_2002}.

\subsection{Generating Synthetic Data}
To create our \numdata{} synthetic data sets, we drew responses as random samples from known populations. We use the term "condition" to refer to combinations of levels from any number of factors, but it is important to note that samples were only drawn for conditions with one level from each factor. Our synthetic data sets varied according to the following four properties:

\begin{itemize}

\item \f{layout}: The number of factors and number of levels per factor in the data set. Values: two factors with two levels each (2 $\times$ 2), two factors with three levels each (3 $\times$ 3), and three factors with two levels each (2 $\times$ 2 $\times$ 2).

\item \f{population distribution}: Type of distribution from which samples in the data set were drawn. Specific distributions (see Table \ref{tab:distributions}) were chosen because they represent data frequently found in HCI studies (e.g., normal, lognormal, exponential), or because they're commonly used in simulation studies in statistics due to their heavy tails \cite{abundis_multiple_2001, blair_comparison_1980} (e.g., Cauchy, \textit{t} with 3 degrees of freedom, double exponential). Note that the mean is a type of location and the standard deviation is a type of scale; for consistency, we use the general terms "location" and "scale."

\begin{table}[h]
    \caption{\f{population distributions} and their parameters.}
    \input{tables/distribution_table.tex}
    \label{tab:distributions}
\end{table}

\item \f{condition sample size}: The number of data points randomly sampled from a population for each condition. Values: 8, 16, 24, 32, and 40, selected because they represented typical sample sizes in HCI.

\item \f{between or within subjects}: In a between-subjects design, each subject contributes one response to the data set, and the number of responses is equal to the number of subjects. In a within-subjects design, each subject contributes one response in each condition, and the number of subjects is equal to the \f{condition sample size}. Values: between and within. Mixed factorial designs were left for future work.

\end{itemize}

Our running example has a 2 $\times$ 2 $\times$ 2 \f{layout}, \f{condition sample size} of 40, lognormal \f{population distribution}, and is within-subjects.
For each of the 3 $\times$ 6 $\times$ 5 $\times$ 2 = 180 combinations of property values, we generated approximately 200 data sets in which all conditions in one data set were sampled from \f{population distributions} with equal locations and approximately another 200 data sets in which all population locations were randomly chosen. Population scale was always equal to 1. We describe our data generating process in four steps:

\smallskip
\noindent \textbf{Step 1. Determine latent location.} We begin by determining a \textit{latent location} for each condition ($\mu^*_{c}$), which will undergo several transformations before being used as parameter values in a \f{population distribution}. When conditions have equal population locations, the latent location is fixed at 0 (Equation \eqref{eq:step1.1a}). Otherwise, its value is sampled from a standard normal distribution (Equation \eqref{eq:step1.1b}). Scale is always equal to 1 in our analyses (Equation \eqref{eq:step1.2}).
\begin{subequations}
\begin{align}
\mu^*_{c} &= 0\label{eq:step1.1a} \\[-4pt]
\intertext{\centering Used when creating data to measure Type I error rate.}
\mu^*_{c} &\sim \mathcal{N}(0,1)\label{eq:step1.1b}
\end{align}\vspace{-.5cm}
\begin{center}Used when creating data to measure statistical power.\end{center}
\end{subequations}
\begin{align}
\sigma_{c} &= 1\label{eq:step1.2}
\end{align}
In Equations \eqref{eq:step1.1a} and \eqref{eq:step1.1b}, $\mu^*_{c}$ is the latent location for condition $c$, and in Equation \eqref{eq:step1.2}, $\sigma_{c}$ is the scale for condition $c$.

\smallskip
\noindent \textbf{Step 2. Add random intercepts per subject.}
When generating within-subjects data, each subject is assigned a unique random offset ($\beta_{s}$) sampled from a normal distribution with mean 0, and standard deviation $SD$ (Equation \eqref{eq:step2.2}), where $SD$ is randomly chosen from $\{0.1, 0.5, 0.9\}$ (Equation \eqref{eq:step2.1}) and is the same value for the entire data set. These values were chosen to represent a reasonable ratio between within-subject variance and between-subject variance. We now update our latent mean notation to ($\mu^*_{c,s}$) to represent the latent mean for each combination of condition and subject, and a subject's random offset is added to all of its associated latent locations (Equation \eqref{eq:step2.3}). For consistency, we use this notation for between-subjects data as well, but with a random per-subject offset of $0$ (Equation \eqref{eq:step2.4}).
\begin{subequations}
\begin{align}
SD &\sim Random(0.1, 0.5, 0.9)\label{eq:step2.1}\\
\beta_s &\sim \mathcal{N}(0, SD)\label{eq:step2.2}\\
\mu^*_{c,s} &= \mu^*_{c} + \beta_{s}\label{eq:step2.3}\\[-3pt]
\intertext{\centering Used when generating within-subjects data}
\mu^*_{c,s} &= \mu^*_{c} + 0\label{eq:step2.4}
\end{align}\vspace{-.5cm}
\begin{center}Used when generating between-subjects data.\end{center}
\end{subequations}
Equations \eqref{eq:step2.1}, \eqref{eq:step2.2}, and \eqref{eq:step2.3} are used for within-sugjects data. In Equation \eqref{eq:step2.2} $\beta_s$ is the random offset for subject $s$, and in Equation \eqref{eq:step2.3} $\beta_s$ is added to all latent locations $\mu^*_{c}$ associated subject $s$, resulting in a new latent location $\mu^*_{c,s}$ for condition $c$ and subject $s$. In Equation \eqref{eq:step2.4} $\mu^*_{c}$ is simply relabeled $\mu^*_{c,s}$ for consistency, however each subject still has the same latent location $\mu^*_{c}$ for condition $c$.

\medskip
\noindent \textbf{Step 3. Transform latent location with inverse link function.}
Latent location is currently expressed as a linear model, but some distributions' parameters must be expressed as a \textit{function} of a linear model. This function, called the \textit{inverse link function} ($g$), transforms the latent location ($\mu^*$) into the appropriate location for the distribution ($\mu$) (Equations \eqref{eq:step3.1}).
\begin{align}
\mu_{c,s} &= g_\mu(\mu^*_{c,s})\label{eq:step3.1}
\end{align}
In Equations \eqref{eq:step3.1}, $g_\mu$ is the location inverse link function and it transforms a latent location $\mu^*_{c,s}$ into a location $\mu_{c,s}$.

All \f{population distributions} (Table \ref{tab:distributions}) use the identity inverse link function (Equations \ref{eq:step3.3}) except for the exponential distribution (Equation \ref{eq:step3.5}).
\begin{subequations}
\begin{align}
g_{\mu,id}(x)&=x\label{eq:step3.3}\\
g_{\mu,exp}(x) &= exp(x)\label{eq:step3.5}
\end{align}
\end{subequations}
In Equations \eqref{eq:step3.3}, $g_{\mu,id}$ is the identity location inverse link function. In Equation \eqref{eq:step3.5}, $g_{\mu,exp}$ is the location inverse link function used by the exponential distribution.

\smallskip
\noindent \textbf{Step 4. Generate data.}
Response $Y_{c,s}$ is sampled from the relevant distribution, represented here by the generic function\\
$Distribution(x,y)$ (Equation \eqref{eq:step4.1}). The exponential distribution only has a single parameter ($rate = 1/location$) and follows Equation
\eqref{eq:step4.2}:
\begin{align}
Y_{c,s} &\sim Distribution(\mu_{c,s}, \sigma_{c,s})\label{eq:step4.1}\\
Y_{c,s} &\sim Exp(1/\mu_{c,s})\label{eq:step4.2}
\end{align}

\noindent \textbf{Example.} We illustrate this process by generating response $Y_{5,2}$ for \textit{condition 5} and \textit{subject 2} in our running example. 
\noindent \textbf{Step 1.} Since our example does not have equal population locations, we use Equations \ref{eq:step1.1b} and \ref{eq:step1.2}.
\begin{align}
\begin{split}
    \mu^*_{5,2} &\sim \mathcal{N}(0, 1)\\
                           & = 0.75
\end{split}\tag{\ref{eq:step1.1b}}\\
\sigma_{5,2} &= 1\tag{\ref{eq:step1.2}}
\end{align}
\noindent \textbf{Step 2.} Our example uses a within-subjects design, so we add per-participant offsets. Note that $SD$ would have already been chosen for \textit{condition 1}. The same value would be used here.
\begin{align}
\begin{split}
    SD &\sim Random(0.1, 0.5, 0.9)\\
       &= 0.5
\end{split}\tag{\ref{eq:step2.1}}\\
\begin{split}
    \beta_2 &\sim \mathcal{N}(0, 0.5)\\
             &= 0.1
\end{split}\tag{\ref{eq:step2.2}}\\
\begin{split}
    \mu^*_{5,2} &= 0.75 + 0.1\\
                           &= 0.85
\end{split}\tag{\ref{eq:step2.3}}
\end{align}

\noindent \textbf{Step 3.} We use the inverse location link function for the lognormal distribution.
\begin{align}
\begin{split}
    \mu_{5,2} &= g_{\mu,id}(0.85) \\
                   &= 0.85
\end{split}\tag{\ref{eq:step3.3}}
\end{align}
\noindent \textbf{Step 4.} Finally, we sample a lognormal distribution with log mean $\mu_{5,2}$ and log standard deviation $\sigma_{5,2}$ to get response $Y_{5,2}$.
\begin{align}
\begin{split}
    Y_{5,2} &\sim Lognormal(0.85, 1)\\
            &= 3.27
\end{split}\tag{\ref{eq:step4.1}}
\end{align}

\subsection{Testing Procedure}
To best explain our testing procedure, we first introduce some definitions:
\begin{itemize}
    \item An \textit{x-factor contrast} is a contrast between two conditions composed of one level each from \textit{x} factors.
    \item \f{contrast size} is the \textit{x} in \textit{x}-factor contrast.
    \item A \textit{design} as a unique combination of a \f{layout}, \f{population distribution}, \f{condition sample size}, \f{between} or \f{within subjects}, and \f{contrast size}. 
   \item A \textit{trial} consists of one contrast test result, and all possible contrasts were conducted. There were:
   \begin{itemize}
        \item 8 trials in a data set with a 2 $\times$ 2 \f{layout}
        \item 42 trials in a data set with a 3 $\times$ 3 \f{layout}
        \item 49 trials in a 2 $\times$ 2 $\times$ 2 \f{layout}.
   \end{itemize}
\end{itemize}

There were \numdata{} total data sets split evenly among designs with each \f{layout} (24,000 data sets each). Thus, there were a total of $24,000 \times (8 + 42 + 49)$ = 2,376,000 trials. There were 1,094 data sets out of 72,000 data sets (1.5\%) with at least one trial for which ART-C did not converge. All of these data sets were within-subjects and were therefore modeled as \textit{linear mixed models} using the \textit{lmer} method in R, and it is common for \textit{lmer} to cause convergence issues. These data sets were removed.

By definition, ART-C is an aligning and ranking procedure followed by a contrast testing method---our validation used a \textit{t}-test. Since we were validating a contrast testing method and not investigating the cause of a significant omnibus result, we did not correct for multiple comparisons. The R programming language was used to generate all data sets, conduct all contrasts, and analyze the results. All R code is included as supplementary material and is available online for replication and extension.\footnote{http://www.doi.org/10.5281/zenodo.4536432}

Following a common approach in statistics literature \cite{abundis_multiple_2001, blair_comparison_1980, li_robustness_nodate, neave_monte_1968, olson_efficacy_2013, peterson_six_2002}, we validated our method on two metrics: Type I error rate and power.

\subsection{Type I Error Rate}
A significance level ($\alpha$) represents the probability of a Type I error (false positive) and is used as a threshold to reject a null hypothesis ($p < \alpha$). Many readers will recognize that typically, alpha is set to .05, although other values may be used. A large-scale simulation shows a method is correct when the proportion of tests in which a true null hypothesis was rejected (\textit{observed Type I error rate}) is close to the significance level. That is, the proportion of tests in which $p$ was less than $\alpha$ should be close to $\alpha$.

For example, 5,516 trials were conducted on data from a 2 $\times$ 2 $\times$ 2 \f{layout}, lognormal \f{population distribution}, \f{condition sample size} 40, \f{within-subjects}, \f{contrast size} three, and no differences between conditions' population locations. Using a significance level of $.05$, ART-C found a significant difference in 265 trials, resulting in a $\frac{265}{5516} = .048$ observed Type I error rate, which is very close to the $\alpha = .05$ significance threshold, indicating the correctness of ART-C for this design.

Each data point in the following results represents the observed Type I error rate for one design. All population locations were set to 0 (Equation \ref{eq:step1.1a}), and thus the null hypothesis that there is no true difference between conditions' population locations is true for all trials. As is common practice in statistics, we include observed Type I error rates for the \textit{t}-test as a baseline comparison \cite{olson_efficacy_2013, neave_monte_1968}. 

Contrasts conducted with ART-C on designs with a combination of \f{contrast size} one, Cauchy \f{population distribution}, and 3 $\times$ 3 or 2 $\times$ 2 $\times$ 2 \f{layout} had inflated observed Type I error rates ($M$ = .373, $SD$ = .076), while \textit{t}-test contrasts did not ($M$ = .025, $SD$ = .004). Those Cauchy designs were considered outliers and were not included in the remainder of our analysis of Type I error rates; we address this further in our discussion.

\begin{table}[H]
    \caption{Mean Type I error rates (and standard deviations) for ART-C and, for comparison, the \textit{t}-test, grouped by \f{contrast size} and \f{layout} over all designs, excluding designs with a Cauchy \f{population distribution}. Values closer to .05 are better, indicating greater correctness. ART-C has comparable Type I error rates to the \textit{t}-test, but as our additional results show, much greater power.}
    \input{tables/Results/type_i_contrast_size_layout}

    \label{tab:type_i_contrast_size_layout}
\end{table}

\begin{figure}[!h]
\centering
\includegraphics[width=\columnwidth]{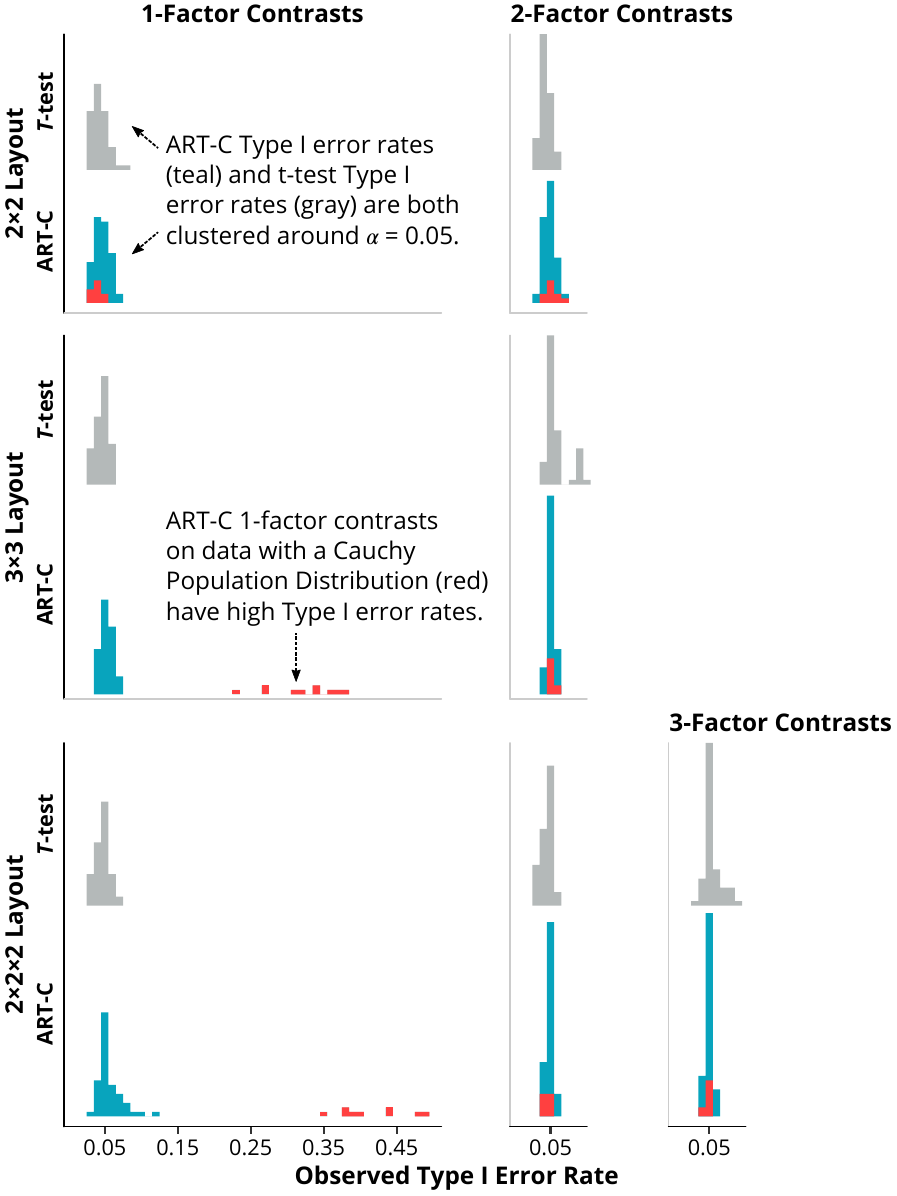}
\caption{ART-C (teal) and \textit{t}-test (gray) observed Type I error rates by \f{contrast size} and \f{layout}. Designs with a Cauchy \f{population distribution} are shown in red. Each point represents observed Type I error rate for one design. Values closer to .05 are better, indicating greater correctness. ART-C has comparable Type I error rates to the \textit{t}-test, but as our additional results show, must greater power.}
\label{fig:type_i_contrast_size_layout}
\end{figure}

Results show observed Type I error rates for contrasts conducted with both ART-C and the \textit{t}-test on remaining designs were clustered around $.05$: ART-C ($M$ = .050, $SD$ = .009) and \textit{t}-test ($M$ = .048, $SD$ = .012), and design properties do not appear to have an effect on observed Type I error rates, confirming the robustness of the ART-C procedure. Observed Type I error rates for all designs are included as supplementary material (see "Type\_I\_all\_designs.csv").
Table \ref{tab:type_i_contrast_size_layout} and Figure \ref{fig:type_i_contrast_size_layout} illustrate both methods' observed Type I error rates, closely clustered around .05, and show egregious Type I error rates for ART-C with a Cauchy distribution.

\subsection{Power}
Statistical power is the probability of rejecting a false null hypothesis (detecting a true difference) given a particular significance level. Observed power is the proportion of tests in which a false null hypothesis was rejected. Unlike for Type I errors, there is no expected value to compare observed power to; instead, we followed common practices in statistics \cite{abundis_multiple_2001, blair_comparison_1980, li_robustness_nodate} and compared to other methods, specifically the \textit{t}-test, Mann-Whitney \textit{U} test for between-subjects designs, and Wilcoxon signed-rank test for within-subjects designs. 

For example, ART-C contrast tests conducted on data with a 2 $\times$ 2 $\times$ 2 \f{layout}, lognormal \f{population distribution}, \f{condition sample size} 40, \f{within-subjects}, \f{contrast size} three, and different population locations for each condition detected a true significant difference in 4408 out of 5824 trials and therefore had $\frac{4408}{5824} = .76$ observed power.

Population locations for each condition were randomly sampled from a standard normal distribution (Equation (\ref{eq:step1.1b})). Although we cannot guarantee these locations were different, there is an infinitely small chance they were the same, and we therefore assume that the null hypothesis of no difference between condition population locations is false. In the following results, a significance level of .05 was used, and each data point represents the observed power of one design.

When averaged over all designs, our results show that ART-C had the highest observed power ($M$ = .598, $SD$ = .143), followed by Mann Whitney \textit{U} test / Wilcoxon signed-rank test ($M$ = .521, $SD$ = .149), and finally the \textit{t}-test ($M$ = .461, $SD$ = .149). Observed powers for all designs are included as supplementary materials (see "power\_all\_designs.csv").

\begin{figure}[H]
\centering
\includegraphics[width=\columnwidth]{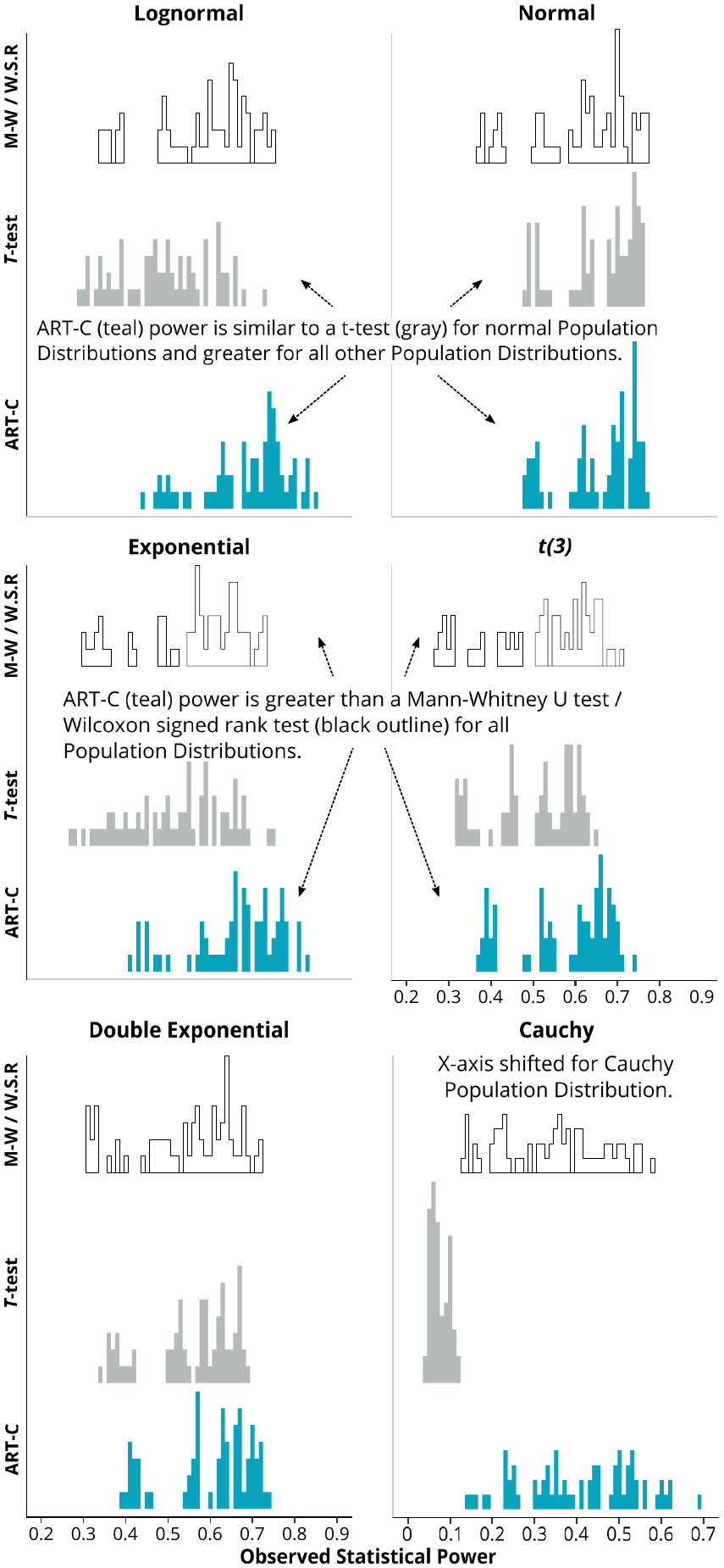}
\caption{Mean statistical power by \f{population distribution} for ART-C (teal), \textit{t}-test (gray), and Mann-Whitney \textit{U} test / Wilcoxon signed rank-test (black outline). Higher values are better indicating greater power. ART-C compares most favorably in all cases except when compared to a \textit{t}-test for normal \f{population distributions}. Each point represents observed statistical power from one design.}
\label{fig:power_distr}
\end{figure}

\f{population distribution} and \f{condition sample size} were the only design properties that had a large impact on observed power. ART-C had higher observed power than the \textit{t}-test for all \f{population distributions} other than the normal distribution, for which it was the same, and had higher observed power than the Mann-Whitney \textit{U} test and Wilcoxon signed-rank test for all \f{population distributions} (Table \ref{tab:power_distribution}, Figure \ref{fig:power_distr}).

\begin{table}[!h]
    \caption{Mean statistical power (and standard deviations) for ART-C, \textit{t}-test, Mann-Whitney \textit{U} test (M-W) / Wilcoxon signed-rank test (W.S.R.), and ART, grouped by \f{population distribution}. Higher values are better, representing more statistical power.}
    \input{tables/Results/power_distribution}
    \label{tab:power_distribution}
\end{table}

\f{condition sample size} affected power in that ART-C had higher observed power than the \textit{t}-test and Mann-Whitney \textit{U} test /\\
Wilcoxon signed-rank test regardless of \f{condition sample size}, but all tests' observed power increased as \f{condition sample size} increased, which is expected.

\subsection{Comparison to ART}
Contrasts conducted with ART-C had lower observed Type I error ($M$ = .067, $SD$ = .072) than contrasts conducted on data aligned-and-ranked using the original ART procedure ($M$ =.121, $SD$ = .174), and higher observed power (ART-C: $M$ = .598, $SD$ = .143 \textit{vs}. ART: $M$ = .511, $SD$ = .182).

When separated by \f{population distribution}, ART-C had lower observed Type I error rates than ART for the lognormal, exponential, and Cauchy \f{population distributions} (Table \ref{tab:ART_type_i_distribution}, Figure \ref{fig:ART_TypeI_Distr}), but ART-C observed Type I error rates were closer to the significance level ($\alpha = .05$) than ART's for all \f{population distributions}, indicating that ART-C is more correct. ART-C also had higher observed power than ART for all \f{population distributions} except Cauchy (Table \ref{tab:power_distribution}, Figure \ref{fig:ART_Power_Distr}).

\begin{table}[!h]
    \caption{Mean Type I error rates (and standard deviations) for ART-C and ART, grouped by \f{population distribution}. Values closer to $\alpha = .05$ are better, indicating greater correctness.}
    \input{tables/Results/ART_type_i_distribution}

    \label{tab:ART_type_i_distribution}
\end{table}

\begin{figure}[!h]
\centering
\includegraphics[width=\columnwidth]{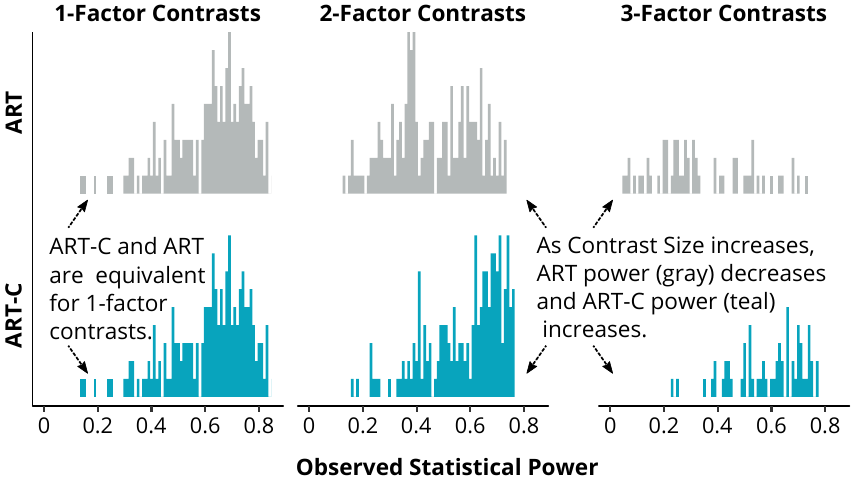}
\caption{ART-C (teal) and ART (gray) observed statistical power by \f{contrast size}. Each point represents observed statistical power for one design. Higher values are better indicating greater power. ART-C power is higher for all \f{contrast sizes} and increases with \f{contrast size} compared to ART which decreases. Both methods are equivalent when conducting single-factor contrasts.}
\label{fig:ART_Power_Contrast_Size}
\end{figure}

\f{contrast size} also had an interesting effect on power. Observed power with ART-C was highest for single-factor contrasts, followed by three-factor contrasts, and then two-factor contrasts, but the differences were small. However, ART's power decreased as \f{contrast size} increased, and the differences were much larger (Table \ref{tab:ART_power_contrast_size}, Figure \ref{fig:ART_Power_Contrast_Size}). Recall that the alignment formulas for ART and ART-C become mathematically equivalent in the single factor case.

\begin{table}[]
    \caption{Mean statistical power (and standard deviations) for ART-C and ART, grouped by \f{contrast size}. Higher values are better, indicating greater power.}
    \input{tables/Results/ART_power_contrast_size}
    \label{tab:ART_power_contrast_size}
\end{table}

Recall in our running example that we conducted multifactor contrasts on levels of factors $A$ and $B$, and contrasts conducted with ART produced a Type I error and a Type II error (see Table \ref{tab:ART_contrasts}, Figure \ref{fig:eg_all_results}), while contrasts conducted with a linear mixed model (LMM) on log-transformed data resulted in correct conclusions (see Table \ref{tab:lmm_contrasts}) that agreed with the ground truth (see Table \ref{tab:eg_true_pop_means}). Contrasts conducted on the same data with ART-C agree with the LMM results and the ground truth in finding a difference between $A1,B1$ and $A2,B2$, and not finding a difference between $A1,B1$ and $A1,B2$ (Table \ref{tab:ART-C_contrasts}).

\begin{table}[!h]
    \caption{Highlighted results of contrasts conducted using ART-C, comparing levels of $A$ and $B$ in our running example. In the top row, a difference was correctly not detected between $A1,B1$ and $A1,B2$ ($p = .6758$), and there is not a true difference. In the bottom row, a difference was detected correctly between $A1,B1$ and $A2,B2$ ($p < .05$), and there \textit{is} a true difference.}
    \input{tables/Running_Eg/ART-C_contrasts}
    \label{tab:ART-C_contrasts}
\end{table}

Thus, taken as a whole, our results show that ART-C has appropriate Type I error rates clustered around $\alpha = .05$, except for data sampled from Cauchy distributions, for which ART-C should not be used. Furthermore, ART-C has high statistical power, outperforming the \textit{t}-test, Mann-Whitney \textit{U} test, Wilcoxon signed-rank test, and ART. These results show that ART-C is a correct and powerful procedure for use within the ART paradigm for conducting contrast tests within or across levels of multiple factors.

%% file: tables/distribution_table.tex
\begin{center}
\begin{tabular}{l|l}
\toprule
\multicolumn{1}{c}{\textbf{Distribution}}   &
\multicolumn{1}{|c}{\textbf{Parameters}}               \\
\midrule
Normal             & Mean, Standard Deviation                 \\
\midrule
Lognormal          & Log Mean, Log Standard Deviation \\
\midrule
Exponential        & Rate                                    \\
\midrule
Cauchy             & Location, Scale                              \\
\midrule
\textit{t}(3)       & Location, Scale  \\
\midrule
Double Exponential & Location, Scale   \\ 
\bottomrule
\end{tabular}
\end{center}

%% file: tables/Results/type_i_contrast_size_layout.tex
\begin{center}
\begin{tabular}{l|l|l|l}
\toprule
\multicolumn{1}{c}{\textbf{\begin{tabular}[c]{@{}c@{}}Contrast\\Size\end{tabular}}} & 
\multicolumn{1}{|c}{\textbf{Layout}} &
\multicolumn{1}{|c}{\textbf{ART-C}} &
\multicolumn{1}{|c}{\textbf{\textit{T}-test}} \\
\midrule
1 & 2$\times$2 & .046 (.011) & .040 (.014) \\ 
\midrule
1 & 3$\times$3 & .053 (.009) & .048 (.007) \\ 
\midrule
1 & 2$\times$2$\times$2 & .057 (.016) & .048 (.009) \\ 
\midrule
2 & 2$\times$2 & .048 (.008) & .042 (.009) \\ 
\midrule
2 & 3$\times$3 & .050 (.004) & .058 (.016) \\ 
\midrule
2 & 2$\times$2$\times$2 & .048 (.005) & .045 (.007) \\ 
\midrule
3 & 2$\times$2$\times$2 & .049 (.004) & .054 (.011) \\ 
\bottomrule
\end{tabular}
\end{center}

%% file: tables/Results/power_distribution.tex

\begin{centering}
\begin{tabular}{l|l|l|l}
\toprule
&
\multicolumn{1}{c}{\textbf{Lognormal}} &
\multicolumn{1}{|c}{\textbf{Normal}} &
\multicolumn{1}{|c}{\textbf{Exponential}}\\
\midrule
ART-C & .69 (.10) & .66 (.09) & .66 (.11) \\ 
\midrule
\textit{T}-test & .46 (.11) & .66 (.09) & .52 (.12) \\ 
\midrule
M-W/ W.S.R. & .58 (.12) & .59 (.12) & .56 (.12) \\ 
\midrule
 ART & .62 (.14) & .49 (.18) & .58 (.16) \\ 
\bottomrule

\multicolumn{4}{}{}\\ 

\toprule
& \multicolumn{1}{|c}{\textbf{\textit{t}(3)}}
& \multicolumn{1}{|c}{\textbf{\begin{tabular}[c]{@{}c@{}}Double\\Exp.\end{tabular}}} & 
\multicolumn{1}{|c}{\textbf{Cauchy}}\\
\midrule
ART-C & .58 (.11) & .60 (.10) & .41 (.13) \\ 
\midrule
\textit{T}-test & .50 (.10) & .56 (.10) & .07 (.02) \\ 
\midrule
M-W/ W.S.R. & .52 (.13) & .54 (.12) & .34 (.12) \\ 
\midrule
ART & .42 (.18) & .44 (.19) & .51 (.16) \\ 

\bottomrule
\end{tabular}
\end{centering}

%% file: tables/Results/ART_type_i_distribution.tex
\begin{center}
\begin{tabular}{l|l|l|l}
    \toprule
    &\multicolumn{1}{c}{\textbf{Lognormal}} &
    \multicolumn{1}{|c}{\textbf{Normal}} &
    \multicolumn{1}{|c}{\textbf{Exponential}}\\
    \midrule
ART-C & .054 (.015) & .049 (.008) & .051 (.007) \\ 
\midrule
  ART & .141 (.096) & .024 (.023) & .065 (.048) \\ 
    \bottomrule
    
    \multicolumn{4}{}{}\\ 
    
    \toprule

    & \multicolumn{1}{c}{\textbf{\textit{t}(3)}}
    & \multicolumn{1}{|c}{\textbf{\begin{tabular}[c]{@{}c@{}}Double\\Exp.\end{tabular}}} & 
    \multicolumn{1}{|c}{\textbf{Cauchy}}\\
    \midrule
ART-C & .049 (.006) & .049 (.007) & .14 (.154) \\ 
\midrule
  ART & .033 (.015) & .026 (.023) & .425 (.209) \\  
    \bottomrule
\end{tabular}
\end{center}

%% file: tables/Results/ART_power_contrast_size.tex
\begin{center}
\begin{tabular}{l|l|l|l}
  \toprule
  & \multicolumn{1}{c}{\textbf{\begin{tabular}[c]{@{}c@{}}1-Factor\\Contrasts\end{tabular}}}
  & \multicolumn{1}{|c}{\textbf{\begin{tabular}[c]{@{}c@{}}2-Factor\\Contrasts\end{tabular}}}
  & \multicolumn{1}{|c}{\textbf{\begin{tabular}[c]{@{}c@{}}3-Factor\\Contrasts\end{tabular}}}\\ 
  \midrule
ART-C & .620 (.150) & .580 (.140) & .590 (.130) \\ 
\midrule
  ART & .620 (.150) & .460 (.150) & .340 (.180) \\ 
   \bottomrule
\end{tabular}
\end{center}

%% file: tables/Running_Eg/ART-C_contrasts.tex
  \begin{center}
\begin{tabular}{lrrrrl}
  \hline
contrast & estimate & SE & df & t.ratio & p.value \\ 
  \hline
A1,B1 - A1,B2 & 1.5 & 3.7 & 273 & 0.4 & 0.6758 \\
  A1,B1 - A2,B2 & -89.9 & 3.7 & 273 & -24.5 & $<$.0001 \\  
   \hline
   \end{tabular}
\end{center}

%% file: content/tools.tex
\section{ARTool.EXE AND R PACKAGE ``ARTOOL''}
To make ART-C widely available to the community, we updated the existing open-source ARTool.exe Windows application (see footnote 2) \cite{wobbrock_aligned_2011} and the ARTool R package (see footnote 1) to include ART-C.

\subsection{ARTool.exe Windows Application}
\begin{figure}[!h]
     \centering
     \begin{subfigure}[b]{\columnwidth}
         \centering
         \includegraphics[width=\textwidth]{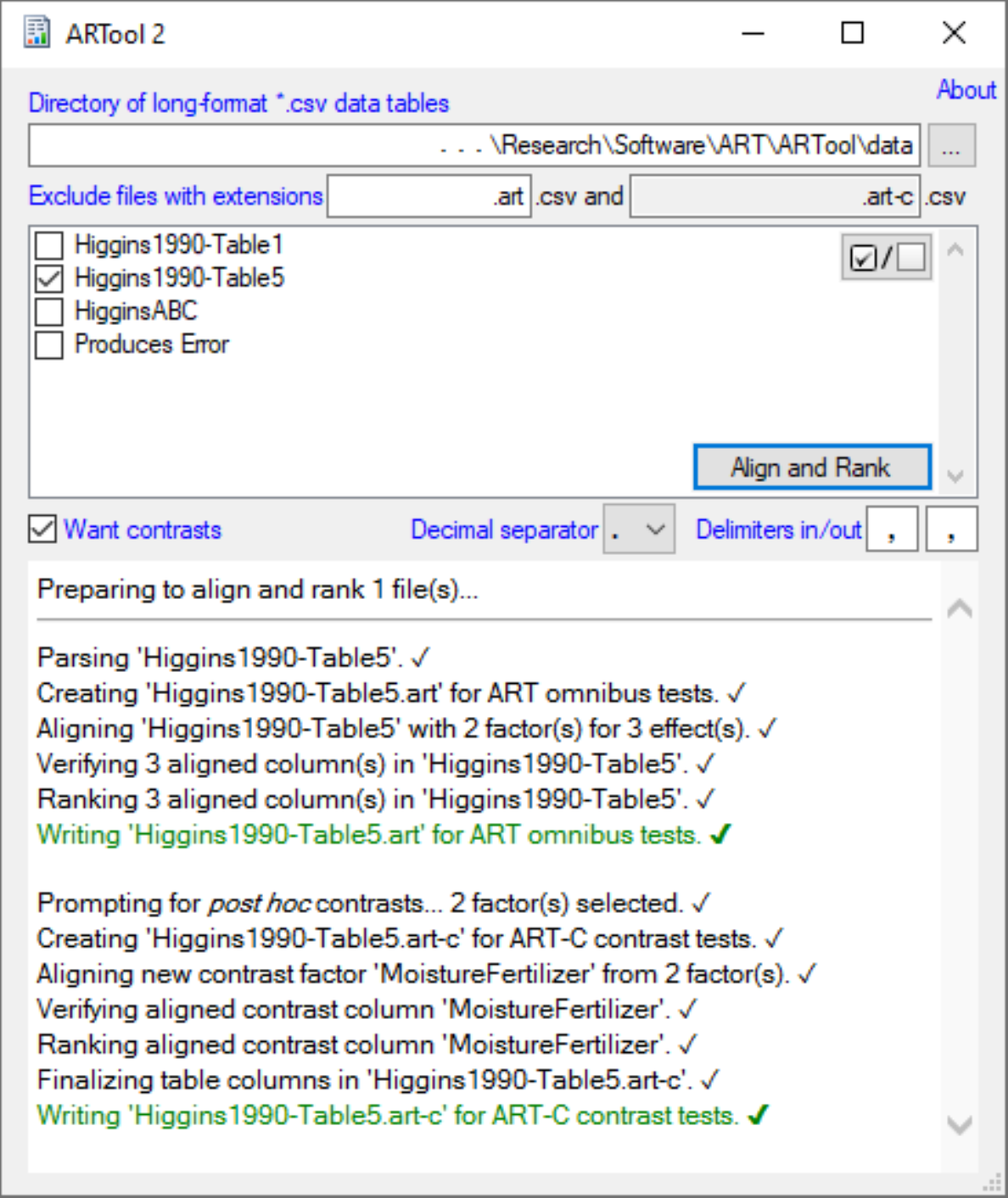}
     \end{subfigure}
     \vfill
     \begin{subfigure}[b]{\columnwidth}
         \centering
         \includegraphics[width=\textwidth]{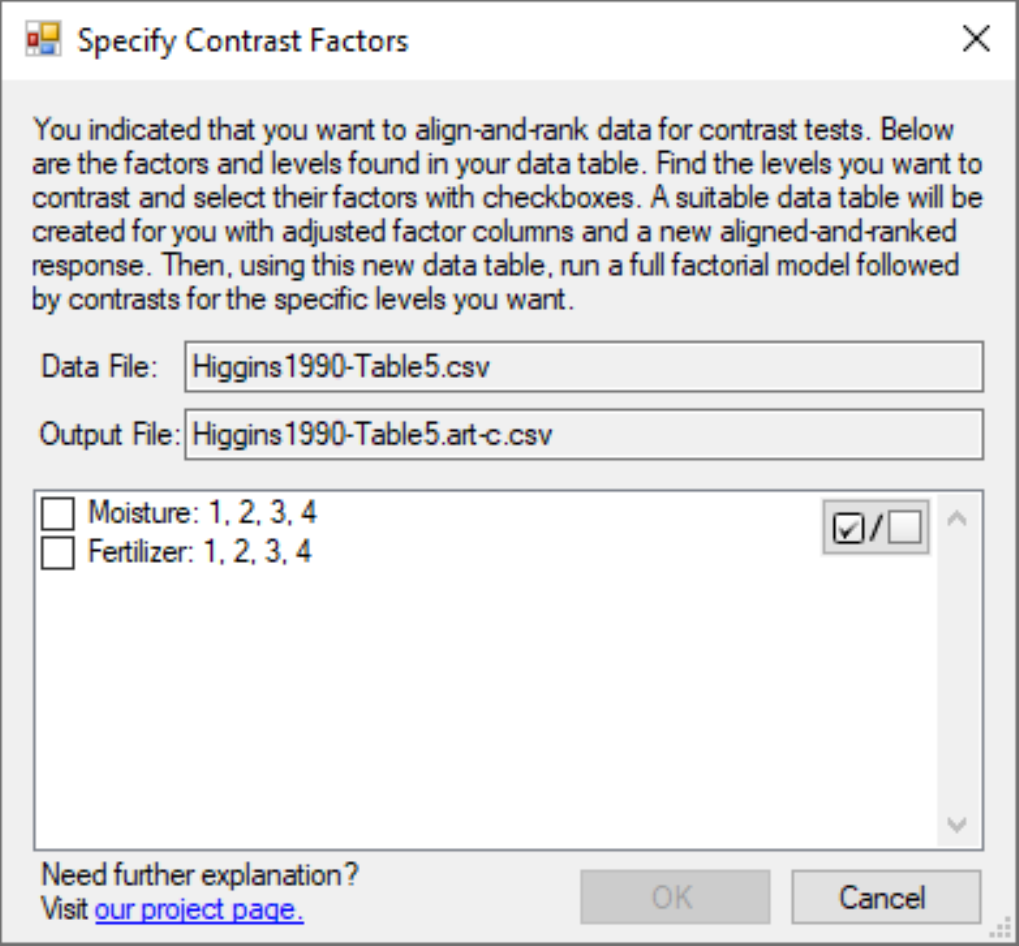}
     \end{subfigure}
        \caption{Top: Our ARTool.exe tool with a “Want contrasts” checkbox. Table 5 from Higgins et al. \cite{higgins_aligned_1990} is being aligned and ranked. Bottom: Our new tool for specifying multifactor contrasts. Two factors, \textit{Moisture} and \textit{Fertilizer}, from the same Table 5, each have levels 1-4. Selecting both factors would allow, e.g., a comparison of (Moisture 2, Fertilizer 3) \textit{vs.} (Moisture 4, Fertilizer 1).}
        \label{fig:windows_artool}
\end{figure}

The ARTool.exe Windows application was released as an open-source tool in 2011 \cite{wobbrock_aligned_2011} to facilitate the aligning and ranking of data for analysis using the ART procedure. We extended this open-source tool to include our ART-C procedure for multifactor contrasts. Users can now indicate that they want contrasts with a checkbox (Figure \ref{fig:windows_artool} (top)), which then offers them a separate dialog box (Figure \ref{fig:windows_artool} (bottom)) from which they can select the factors whose levels are involved in their desired contrast test. ARTool then uses the ART-C procedure to produce aligned and ranked output suitable for analysis. See footnote 3 for the link to our updated ARTool Windows application.


\subsection{R Package ``ARTool''}
The current ``ARTool'' R package makes it easy to conduct nonparametric tests of main effects and interactions using the ART procedure. A single function call aligns and ranks data for each fixed effect in a formula \textit{f} provided by the user. The result is an ART model \textit{m} that keeps a copy of formula \textit{f} and the data. Given \textit{m}, another function in ARTool, \textit{anova}, runs multiple ANOVAs behind the scenes, one for each fixed effect in \textit{f}, and returns the results of each test. In this work, we have added a new function, \textit{art.con}, that uses our ART-C procedure to conduct multifactor contrast tests. Given the same model \textit{m} and a contrast formula $f_c$, the ART-C procedure is used to align and rank the data saved in \textit{m} for the contrasts specified in $f_c$. It then parses the formula \textit{f} saved in \textit{m}, conducts the contrasts, and returns the results. 

\begin{figure}[!h]
\centering
\includegraphics[width=\columnwidth]{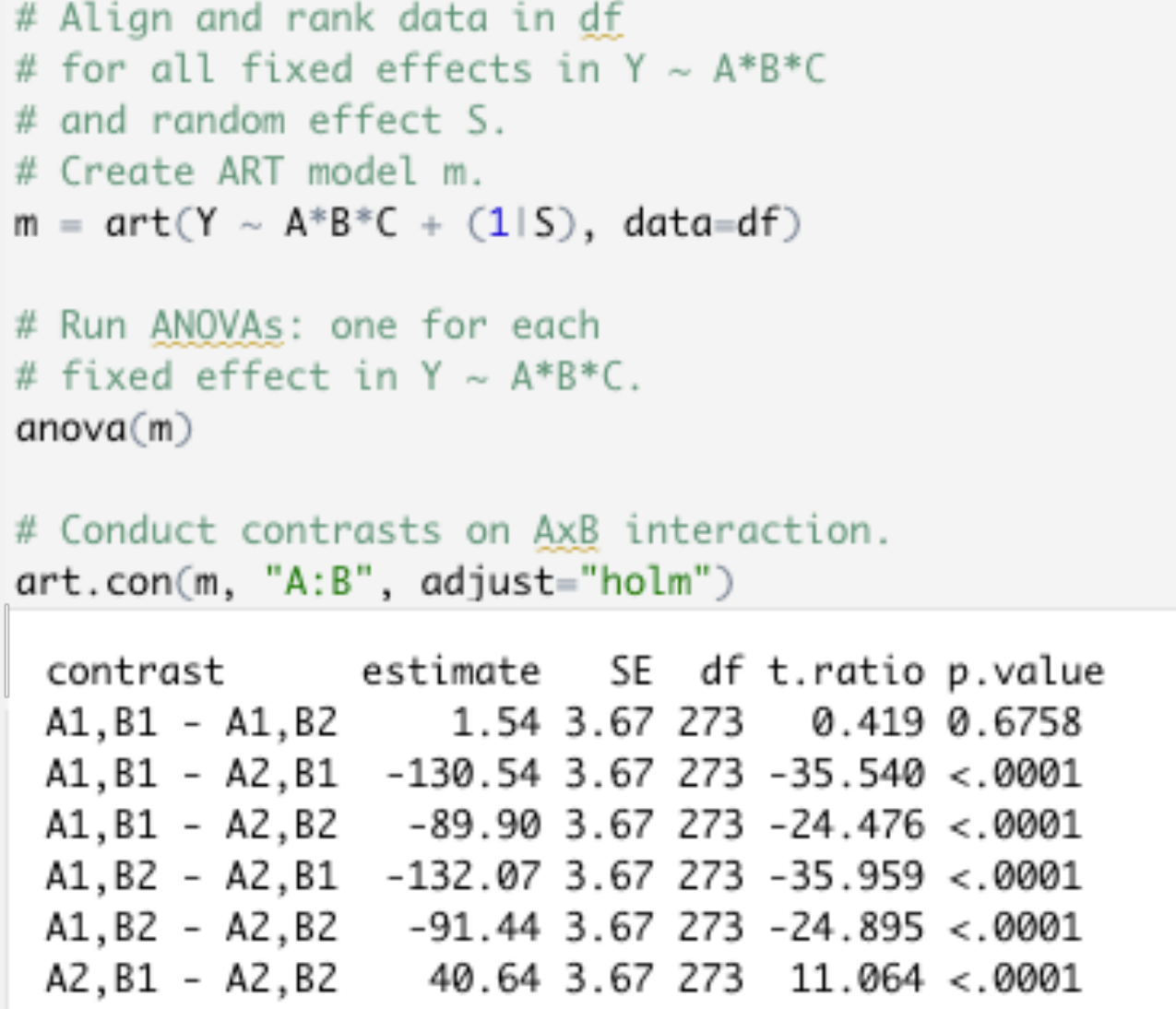}
\caption{Screenshot using ART-C to conduct $A \times B$ contrasts in our running example in R. The \textit{anova} call first would produce omnibus test results for any $A$, $B$, and $C$ main effects and interactions; if, for example, the $A \times B$ interaction were statistically significant, \textit{art.con} could be used to conduct \textit{post hoc} pairwise comparisons as shown here.}
\label{fig:eg_r_screenshot}
\end{figure}

In our running example, we first conducted $A \times B$ contrasts with ART, which, of course, is incorrect given ART's propensity for Type I errors. Now, we can correctly use ART-C to perform these contrasts. Figure \ref{fig:eg_r_screenshot} shows how we would use ART-C to conduct contrasts correctly.

See footnote 1 for the link to our updated R package and footnote 2 for the link to the code.

%% file: content/discussion.tex
\section{Discussion}
Our results showed that ART-C's Type I error rate is typically clustered around $\alpha=.05$, offering strong evidence for its correctness. In addition to having more power than a \textit{t}-test for all non-normal distributions and more power than a Mann-Whitney \textit{U} test and Wilcoxon signed-rank test for all distributions, it is worth noting that the increase in power achieved by using ART-C is largest for data drawn from lognormal and exponential distributions. This finding is particularly meaningful because the lognormal and exponential distributions were included due to their frequent appearance in HCI.

That said, our results showed that single-factor ART-C contrasts conducted on data drawn from a Cauchy-distributed population had high observed Type I error rates. This is not unique to ART-C; the Cauchy distribution is known to be ``pathological'' and many well-known statistics concepts do not hold on Cauchy-distributed data (e.g., the Central Limit Theorem \cite{krishnamoorthy2016handbook}). This situation occurs because Cauchy distributions have tails that are so fat that neither their mean nor variance is well defined. In practice, this concern can arise in data with extreme outliers. Thus, we encourage users to avoid using ART-C if they have theoretical reasons to suspect the data is drawn from a Cauchy-distributed population or if the data has extreme outliers.

In HCI, \nonparam{} tests are typically used as a catch-all when parametric tests are not appropriate. The particular Cauchy result above illuminates that this practice can be problematic. In fact, ART-C is mathematically equivalent to ART in the single-factor case, and ART was thought to be appropriate for single-factor contrasts, but would also be ill-suited for the Cauchy case. A disclaimer to not use a method to analyze data with a particular residual distribution is not useful unless researchers investigate experimental data distributions beyond checking for normality. The American Psychological Association's Taskforce on Statistical Inference encourages researchers to take a closer look at their data:

\begin{quote}
As soon as you have collected your data, before you
compute \emph{any} statistics, \emph{look at your data}. Data screening is
not data snooping. It is not an opportunity to discard data or
change values to favor your hypotheses. However, if you
assess hypotheses without examining your data, you risk
publishing nonsense. \cite{wilkinson1999statistical} (\emph{emphasis in original})
\end{quote}

Even nonparametric tests are subject to assumptions. There are many tried-and-tested visualizations for model diagnostics that can be applied to assess assumptions relevant to ART: quantile-quantile (\textsc{qq}) plots~\cite{hoaglin2006using}, for example, allow one to check for the presence of fat tails in the distribution of residuals (i.e., \emph{excess kurtosis}, which in extreme cases could indicate the presence of Cauchy-distributed data). Modern visualizations like worm plots~\cite{buuren2001worm} can make it even easier to diagnose fat tails. The point, though, is that there is no all-encompassing solution in statistical analysis: model fit and assumptions cannot be assumed and must be checked, and nonparametric approaches are no exception to this rule.

%% file: content/limitations_and_future_work.tex
\section{Limitations and Future Work}
There are infinitely many combinations of layouts, population distributions, and condition sample sizes one could examine in a study like ours, but we could only analyze a finite amount of data and had to be selective. These decisions were carefully made, considering the needs of the HCI community and statistical norms, but they were certainly not exhaustive.

Our validation only investigated data in which all conditions' populations had the same location or all conditions' populations had different locations. Additionally, even when parameter values were varied, conditions in the same data set were always drawn from the same distribution. Data in which there are differences between some conditions and not others arises frequently in HCI, but we chose this validation process because it is commonly used in statistics \cite{abundis_multiple_2001,blair_comparison_1980,li_robustness_nodate,sawilowsky_more_1992}.

We included models with random intercepts, which represented the impact each subject had on the response. However, we did not include models with random slopes, which allow, for some types of responses, better fitting models where subjects' responses vary differentially across another variable (e.g., time). Although random slope models would certainly be valuable, fixed effects models and models with random intercepts are used  more frequently in HCI, so we chose to focus our validation on such models, leaving other models for future work.

ART-C is defined as an alignment procedure, followed by a ranking procedure, and then a contrast test. We chose to use the \textit{t}-test as the contrast test in our analyses because it is the most familiar to the HCI community and therefore how we anticipate most researchers will use our new method. Still, it would be worthwhile to see how ART-C performs when a different contrast test is used.

In addition to our extension to the ARTool R package and ARTool.exe Windows application, we envision a platform-agnostic tool that does not require programming experience, and even an ART and ART-C package for other common statistics software (e.g., SAS, SPSS). With the algorithmic and validation work we have done here, it should be relatively straightforward to create additional add-ons for common statistical packages.

We leave addressing our limitations to future work. To facilitate this, we have published the R code used to generate synthetic data, conduct all contrasts, and log the results (see footnote 5), as well as our analysis code (also footnote 5). The ``ARTool'' R package (see footnotes 1 and 2) and ARTool.exe application (see footnote 3), including our new additions to them, are open source.

%% file: content/conclusion.tex
\section{Conclusion}
The ART method has enabled anyone familiar with an ANOVA to conduct \nonparam{} analyses on data from factorial experiments and detect main effects and interactions \cite{wobbrock_aligned_2011}, but we have shown that it inflates Type I error rates and has low statistical power when used to conduct \multifactor{} contrasts. To remedy this, we have developed, validated, and presented the ART-C procedure for aligning-and-ranking data for \nonparam{} \multifactor{} contrasts within the ART paradigm, giving researchers a technique and tools to analyze data nonparametrically from factorial experiments. We have validated our method's correctness and statistical power on 72,000 synthetic data sets whose properties represent data commonly found within HCI and statistics. Our results show that ART-C does not inflate Type I error, and has higher statistical power than a \textit{t}-test, Mann-Whitney \textit{U} test, Wilcoxon signed-rank test, and ART. To facilitate the widespread use of ART-C, we have added it to the existing ``ARTool'' R package and ARTool.exe Windows application. 

It is our hope that the simplicity of conducting multifactor contrasts with ART-C will impact the HCI community by enabling researchers to easily investigate their nonconforming \multifactor{} data at a level of granularity that previously required statistical expertise and a departure from the ART paradigm. Owing to ART's evident popularity, we believe ART-C can be immediately useful to many researchers in HCI and beyond.

%% file: main.bbl

\begin{thebibliography}{46}


\ifx \showCODEN    \undefined \def \showCODEN     #1{\unskip}     \fi
\ifx \showDOI      \undefined \def \showDOI       #1{#1}\fi
\ifx \showISBNx    \undefined \def \showISBNx     #1{\unskip}     \fi
\ifx \showISBNxiii \undefined \def \showISBNxiii  #1{\unskip}     \fi
\ifx \showISSN     \undefined \def \showISSN      #1{\unskip}     \fi
\ifx \showLCCN     \undefined \def \showLCCN      #1{\unskip}     \fi
\ifx \shownote     \undefined \def \shownote      #1{#1}          \fi
\ifx \showarticletitle \undefined \def \showarticletitle #1{#1}   \fi
\ifx \showURL      \undefined \def \showURL       {\relax}        \fi
\providecommand\bibfield[2]{#2}
\providecommand\bibinfo[2]{#2}
\providecommand\natexlab[1]{#1}
\providecommand\showeprint[2][]{arXiv:#2}

\bibitem[\protect\citeauthoryear{Abundis}{Abundis}{2001}]%
        {abundis_multiple_2001}
\bibfield{author}{\bibinfo{person}{Marisela Abundis}.}
  \bibinfo{year}{2001}\natexlab{}.
\newblock \emph{\bibinfo{title}{Multiple comparison procedures in factorial
  designs using the aligned rank transformation}}.
\newblock Thesis. \bibinfo{school}{Texas Tech University}.
\newblock
\urldef\tempurl%
\url{https://ttu-ir.tdl.org/handle/2346/22569}
\showURL{%
\tempurl}


\bibitem[\protect\citeauthoryear{Amershi, Fogarty, and Weld}{Amershi
  et~al\mbox{.}}{2012}]%
        {amershi_regroup_2012}
\bibfield{author}{\bibinfo{person}{Saleema Amershi}, \bibinfo{person}{James
  Fogarty}, {and} \bibinfo{person}{Daniel Weld}.}
  \bibinfo{year}{2012}\natexlab{}.
\newblock \showarticletitle{Regroup: interactive machine learning for on-demand
  group creation in social networks}. In \bibinfo{booktitle}{\emph{Proceedings
  of the 2012 {ACM} annual conference on {Human} {Factors} in {Computing}
  {Systems} - {CHI} '12}}. \bibinfo{publisher}{ACM Press},
  \bibinfo{address}{Austin, Texas, USA}, \bibinfo{pages}{21}.
\newblock
\showISBNx{978-1-4503-1015-4}
\urldef\tempurl%
\url{https://doi.org/10.1145/2207676.2207680}
\showDOI{\tempurl}


\bibitem[\protect\citeauthoryear{Azenkot, Rector, Ladner, and Wobbrock}{Azenkot
  et~al\mbox{.}}{2012}]%
        {azenkot_passchords_2012}
\bibfield{author}{\bibinfo{person}{Shiri Azenkot}, \bibinfo{person}{Kyle
  Rector}, \bibinfo{person}{Richard Ladner}, {and} \bibinfo{person}{Jacob
  Wobbrock}.} \bibinfo{year}{2012}\natexlab{}.
\newblock \showarticletitle{{PassChords}: secure multi-touch authentication for
  blind people}. In \bibinfo{booktitle}{\emph{Proceedings of the 14th
  international {ACM} {SIGACCESS} conference on {Computers} and accessibility -
  {ASSETS} '12}}. \bibinfo{publisher}{ACM Press}, \bibinfo{address}{Boulder,
  Colorado, USA}, \bibinfo{pages}{159}.
\newblock
\showISBNx{978-1-4503-1321-6}
\urldef\tempurl%
\url{https://doi.org/10.1145/2384916.2384945}
\showDOI{\tempurl}


\bibitem[\protect\citeauthoryear{Barefield and Mansouri}{Barefield and
  Mansouri}{2001}]%
        {barefield_empirical_2001}
\bibfield{author}{\bibinfo{person}{Eric Barefield} {and} \bibinfo{person}{H.
  Mansouri}.} \bibinfo{year}{2001}\natexlab{}.
\newblock \showarticletitle{An empirical study of nonparametric multiple
  comparison procedures in randomized blocks}.
\newblock \bibinfo{journal}{\emph{Journal of Nonparametric Statistics}}
  \bibinfo{volume}{13}, \bibinfo{number}{4} (\bibinfo{date}{Jan.}
  \bibinfo{year}{2001}), \bibinfo{pages}{591--604}.
\newblock
\showISSN{1048-5252, 1029-0311}
\urldef\tempurl%
\url{https://doi.org/10.1080/10485250108832867}
\showDOI{\tempurl}


\bibitem[\protect\citeauthoryear{Blair and Higgins}{Blair and Higgins}{1980}]%
        {blair_comparison_1980}
\bibfield{author}{\bibinfo{person}{R.~Clifford Blair} {and}
  \bibinfo{person}{James~J. Higgins}.} \bibinfo{year}{1980}\natexlab{}.
\newblock \showarticletitle{A {Comparison} of the {Power} of {Wilcoxon}'s
  {Rank}-{Sum} {Statistic} to {That} of {Student}'s t {Statistic} under
  {Various} {Nonnormal} {Distributions}}.
\newblock \bibinfo{journal}{\emph{Journal of Educational Statistics}}
  \bibinfo{volume}{5}, \bibinfo{number}{4} (\bibinfo{year}{1980}),
  \bibinfo{pages}{309}.
\newblock
\showISSN{03629791}
\urldef\tempurl%
\url{https://doi.org/10.2307/1164905}
\showDOI{\tempurl}


\bibitem[\protect\citeauthoryear{Buuren and Fredriks}{Buuren and
  Fredriks}{2001}]%
        {buuren2001worm}
\bibfield{author}{\bibinfo{person}{Stef~van Buuren} {and}
  \bibinfo{person}{Miranda Fredriks}.} \bibinfo{year}{2001}\natexlab{}.
\newblock \showarticletitle{Worm plot: a simple diagnostic device for modelling
  growth reference curves}.
\newblock \bibinfo{journal}{\emph{Statistics in medicine}}
  \bibinfo{volume}{20}, \bibinfo{number}{8} (\bibinfo{year}{2001}),
  \bibinfo{pages}{1259--1277}.
\newblock


\bibitem[\protect\citeauthoryear{Ciavardelli, Rossi, Barcaroli, Volpe,
  Consalvo, Zucchelli, De~Cola, Scavo, Carollo, D'Agostino, Forlì, D'Aguanno,
  Todaro, Stassi, Di~Ilio, De~Laurenzi, and Urbani}{Ciavardelli
  et~al\mbox{.}}{2014}]%
        {ciavardelli_breast_2014}
\bibfield{author}{\bibinfo{person}{D. Ciavardelli}, \bibinfo{person}{C. Rossi},
  \bibinfo{person}{D. Barcaroli}, \bibinfo{person}{S. Volpe},
  \bibinfo{person}{A. Consalvo}, \bibinfo{person}{M. Zucchelli},
  \bibinfo{person}{A. De~Cola}, \bibinfo{person}{E. Scavo}, \bibinfo{person}{R.
  Carollo}, \bibinfo{person}{D. D'Agostino}, \bibinfo{person}{F. Forlì},
  \bibinfo{person}{S. D'Aguanno}, \bibinfo{person}{M. Todaro},
  \bibinfo{person}{G. Stassi}, \bibinfo{person}{C. Di~Ilio},
  \bibinfo{person}{V. De~Laurenzi}, {and} \bibinfo{person}{A. Urbani}.}
  \bibinfo{year}{2014}\natexlab{}.
\newblock \showarticletitle{Breast cancer stem cells rely on fermentative
  glycolysis and are sensitive to 2-deoxyglucose treatment}.
\newblock \bibinfo{journal}{\emph{Cell Death \& Disease}} \bibinfo{volume}{5},
  \bibinfo{number}{7} (\bibinfo{date}{July} \bibinfo{year}{2014}),
  \bibinfo{pages}{e1336--e1336}.
\newblock
\showISSN{2041-4889}
\urldef\tempurl%
\url{https://doi.org/10.1038/cddis.2014.285}
\showDOI{\tempurl}
\newblock
\shownote{Number: 7 Publisher: Nature Publishing Group.}


\bibitem[\protect\citeauthoryear{Conover and Iman}{Conover and Iman}{1981}]%
        {conover_rank_1981}
\bibfield{author}{\bibinfo{person}{W.~J. Conover} {and}
  \bibinfo{person}{Ronald~L. Iman}.} \bibinfo{year}{1981}\natexlab{}.
\newblock \showarticletitle{Rank {Transformations} as a {Bridge} {Between}
  {Parametric} and {Nonparametric} {Statistics}}.
\newblock \bibinfo{journal}{\emph{The American Statistician}}
  \bibinfo{volume}{35}, \bibinfo{number}{3} (\bibinfo{year}{1981}),
  \bibinfo{pages}{124--129}.
\newblock
\showISSN{00031305}
\urldef\tempurl%
\url{http://www.jstor.org/stable/2683975}
\showURL{%
\tempurl}


\bibitem[\protect\citeauthoryear{Feilich and Lauder}{Feilich and
  Lauder}{2015}]%
        {feilich_passive_2015}
\bibfield{author}{\bibinfo{person}{Kara~L. Feilich} {and}
  \bibinfo{person}{George~V. Lauder}.} \bibinfo{year}{2015}\natexlab{}.
\newblock \showarticletitle{Passive mechanical models of fish caudal fins:
  effects of shape and stiffness on self-propulsion}.
\newblock \bibinfo{journal}{\emph{Bioinspiration \& Biomimetics}}
  \bibinfo{volume}{10}, \bibinfo{number}{3} (\bibinfo{date}{April}
  \bibinfo{year}{2015}), \bibinfo{pages}{036002}.
\newblock
\showISSN{1748-3190}
\urldef\tempurl%
\url{https://doi.org/10.1088/1748-3190/10/3/036002}
\showDOI{\tempurl}
\newblock
\shownote{Publisher: IOP Publishing.}


\bibitem[\protect\citeauthoryear{Frederick}{Frederick}{1999}]%
        {frederick_fixed-_1999}
\bibfield{author}{\bibinfo{person}{Brigitte~N. Frederick}.}
  \bibinfo{year}{1999}\natexlab{}.
\newblock \bibinfo{booktitle}{\emph{Fixed-, {Random}-, and {Mixed}-{Effects}
  {ANOVA} {Models}: {A} {User}-{Friendly} {Guide} for {Increasing} the
  {Generalizability} of {ANOVA} {Results}}}.
\newblock
\urldef\tempurl%
\url{https://eric.ed.gov/?id=ED426098}
\showURL{%
\tempurl}


\bibitem[\protect\citeauthoryear{Friedman}{Friedman}{1937}]%
        {friedman_use_1937}
\bibfield{author}{\bibinfo{person}{Milton Friedman}.}
  \bibinfo{year}{1937}\natexlab{}.
\newblock \showarticletitle{The {Use} of {Ranks} to {Avoid} the {Assumption} of
  {Normality} {Implicit} in the {Analysis} of {Variance}}.
\newblock \bibinfo{journal}{\emph{J. Amer. Statist. Assoc.}}
  \bibinfo{volume}{32}, \bibinfo{number}{200} (\bibinfo{year}{1937}),
  \bibinfo{pages}{675--701}.
\newblock
\showISSN{0162-1459}
\urldef\tempurl%
\url{https://doi.org/10.2307/2279372}
\showDOI{\tempurl}
\newblock
\shownote{Publisher: [American Statistical Association, Taylor \& Francis,
  Ltd.].}


\bibitem[\protect\citeauthoryear{Gaspar, Lourenço, Pereira, Azevedo,
  Roncon-Albuquerque, Marques, and Leite-Moreira}{Gaspar et~al\mbox{.}}{2018}]%
        {gaspar_randomized_2018}
\bibfield{author}{\bibinfo{person}{António Gaspar}, \bibinfo{person}{André~P.
  Lourenço}, \bibinfo{person}{Miguel~Álvares Pereira}, \bibinfo{person}{Pedro
  Azevedo}, \bibinfo{person}{Roberto Roncon-Albuquerque},
  \bibinfo{person}{Jorge Marques}, {and} \bibinfo{person}{Adelino~F.
  Leite-Moreira}.} \bibinfo{year}{2018}\natexlab{}.
\newblock \showarticletitle{Randomized controlled trial of remote ischaemic
  conditioning in {ST}-elevation myocardial infarction as adjuvant to primary
  angioplasty ({RIC}-{STEMI})}.
\newblock \bibinfo{journal}{\emph{Basic Research in Cardiology}}
  \bibinfo{volume}{113}, \bibinfo{number}{3} (\bibinfo{date}{March}
  \bibinfo{year}{2018}), \bibinfo{pages}{14}.
\newblock
\showISSN{1435-1803}
\urldef\tempurl%
\url{https://doi.org/10.1007/s00395-018-0672-3}
\showDOI{\tempurl}


\bibitem[\protect\citeauthoryear{Gugenheimer, Stemasov, Frommel, and
  Rukzio}{Gugenheimer et~al\mbox{.}}{2017}]%
        {gugenheimer_sharevr_2017}
\bibfield{author}{\bibinfo{person}{Jan Gugenheimer}, \bibinfo{person}{Evgeny
  Stemasov}, \bibinfo{person}{Julian Frommel}, {and} \bibinfo{person}{Enrico
  Rukzio}.} \bibinfo{year}{2017}\natexlab{}.
\newblock \showarticletitle{{ShareVR}: {Enabling} {Co}-{Located} {Experiences}
  for {Virtual} {Reality} between {HMD} and {Non}-{HMD} {Users}}. In
  \bibinfo{booktitle}{\emph{Proceedings of the 2017 {CHI} {Conference} on
  {Human} {Factors} in {Computing} {Systems}}}. \bibinfo{publisher}{ACM},
  \bibinfo{address}{Denver Colorado USA}, \bibinfo{pages}{4021--4033}.
\newblock
\showISBNx{978-1-4503-4655-9}
\urldef\tempurl%
\url{https://doi.org/10.1145/3025453.3025683}
\showDOI{\tempurl}


\bibitem[\protect\citeauthoryear{Hamdan, Wagner, Voelker, Steimle, and
  Borchers}{Hamdan et~al\mbox{.}}{2019}]%
        {hamdan_springlets_2019}
\bibfield{author}{\bibinfo{person}{Nur Al-huda Hamdan}, \bibinfo{person}{Adrian
  Wagner}, \bibinfo{person}{Simon Voelker}, \bibinfo{person}{Jürgen Steimle},
  {and} \bibinfo{person}{Jan Borchers}.} \bibinfo{year}{2019}\natexlab{}.
\newblock \showarticletitle{Springlets: {Expressive}, {Flexible} and {Silent}
  {On}-{Skin} {Tactile} {Interfaces}}. In \bibinfo{booktitle}{\emph{Proceedings
  of the 2019 {CHI} {Conference} on {Human} {Factors} in {Computing} {Systems}
  - {CHI} '19}}. \bibinfo{publisher}{ACM Press}, \bibinfo{address}{Glasgow,
  Scotland Uk}, \bibinfo{pages}{1--14}.
\newblock
\showISBNx{978-1-4503-5970-2}
\urldef\tempurl%
\url{https://doi.org/10.1145/3290605.3300718}
\showDOI{\tempurl}


\bibitem[\protect\citeauthoryear{Higgins, Blair, and Tashtoush}{Higgins
  et~al\mbox{.}}{1990}]%
        {higgins_aligned_1990}
\bibfield{author}{\bibinfo{person}{James~J. Higgins},
  \bibinfo{person}{R.~Clifford Blair}, {and} \bibinfo{person}{Suleiman
  Tashtoush}.} \bibinfo{year}{1990}\natexlab{}.
\newblock \showarticletitle{{THE} {ALIGNED} {RANK} {TRANSFORM} {PROCEDURE}}.
\newblock \bibinfo{journal}{\emph{Conference on Applied Statistics in
  Agriculture}} (\bibinfo{date}{April} \bibinfo{year}{1990}).
\newblock
\showISSN{2475-7772}
\urldef\tempurl%
\url{https://doi.org/10.4148/2475-7772.1443}
\showDOI{\tempurl}


\bibitem[\protect\citeauthoryear{Higgins and Tashtoush}{Higgins and
  Tashtoush}{1994}]%
        {higgins_1994_aligned}
\bibfield{author}{\bibinfo{person}{James~J Higgins} {and}
  \bibinfo{person}{Suleiman Tashtoush}.} \bibinfo{year}{1994}\natexlab{}.
\newblock \showarticletitle{An aligned rank transform test for interaction}.
\newblock \bibinfo{journal}{\emph{Nonlinear World}} \bibinfo{volume}{1},
  \bibinfo{number}{2} (\bibinfo{year}{1994}), \bibinfo{pages}{201--211}.
\newblock


\bibitem[\protect\citeauthoryear{Hoaglin}{Hoaglin}{2006}]%
        {hoaglin2006using}
\bibfield{author}{\bibinfo{person}{David~C Hoaglin}.}
  \bibinfo{year}{2006}\natexlab{}.
\newblock \showarticletitle{Using quantiles to study shape}.
\newblock \bibinfo{journal}{\emph{Exploring data tables, trends, and shapes}}
  (\bibinfo{year}{2006}), \bibinfo{pages}{417--460}.
\newblock


\bibitem[\protect\citeauthoryear{Hodges and Lehmann}{Hodges and
  Lehmann}{1962}]%
        {hodges_rank_1962}
\bibfield{author}{\bibinfo{person}{J.~L. Hodges} {and} \bibinfo{person}{E.~L.
  Lehmann}.} \bibinfo{year}{1962}\natexlab{}.
\newblock \showarticletitle{Rank {Methods} for {Combination} of {Independent}
  {Experiments} in {Analysis} of {Variance}}.
\newblock \bibinfo{journal}{\emph{The Annals of Mathematical Statistics}}
  \bibinfo{volume}{33}, \bibinfo{number}{2} (\bibinfo{date}{June}
  \bibinfo{year}{1962}), \bibinfo{pages}{482--497}.
\newblock
\showISSN{0003-4851, 2168-8990}
\urldef\tempurl%
\url{https://doi.org/10.1214/aoms/1177704575}
\showDOI{\tempurl}


\bibitem[\protect\citeauthoryear{Holm}{Holm}{1979}]%
        {holm1979simple}
\bibfield{author}{\bibinfo{person}{Sture Holm}.}
  \bibinfo{year}{1979}\natexlab{}.
\newblock \showarticletitle{A simple sequentially rejective multiple test
  procedure}.
\newblock \bibinfo{journal}{\emph{Scandinavian journal of statistics}}
  (\bibinfo{year}{1979}), \bibinfo{pages}{65--70}.
\newblock


\bibitem[\protect\citeauthoryear{Jun, Daum, Roesch, Chasins, Berger, Just, and
  Reinecke}{Jun et~al\mbox{.}}{2019}]%
        {jun_tea_2019}
\bibfield{author}{\bibinfo{person}{Eunice Jun}, \bibinfo{person}{Maureen Daum},
  \bibinfo{person}{Jared Roesch}, \bibinfo{person}{Sarah Chasins},
  \bibinfo{person}{Emery Berger}, \bibinfo{person}{Rene Just}, {and}
  \bibinfo{person}{Katharina Reinecke}.} \bibinfo{year}{2019}\natexlab{}.
\newblock \showarticletitle{Tea: {A} {High}-level {Language} and {Runtime}
  {System} for {Automating} {Statistical} {Analysis}}. In
  \bibinfo{booktitle}{\emph{Proceedings of the 32nd {Annual} {ACM} {Symposium}
  on {User} {Interface} {Software} and {Technology}}}.
  \bibinfo{publisher}{ACM}, \bibinfo{address}{New Orleans LA USA},
  \bibinfo{pages}{591--603}.
\newblock
\showISBNx{978-1-4503-6816-2}
\urldef\tempurl%
\url{https://doi.org/10.1145/3332165.3347940}
\showDOI{\tempurl}


\bibitem[\protect\citeauthoryear{Kane, Morris, Perkins, Wigdor, Ladner, and
  Wobbrock}{Kane et~al\mbox{.}}{2011}]%
        {kane_access_2011}
\bibfield{author}{\bibinfo{person}{Shaun~K. Kane},
  \bibinfo{person}{Meredith~Ringel Morris}, \bibinfo{person}{Annuska~Z.
  Perkins}, \bibinfo{person}{Daniel Wigdor}, \bibinfo{person}{Richard~E.
  Ladner}, {and} \bibinfo{person}{Jacob~O. Wobbrock}.}
  \bibinfo{year}{2011}\natexlab{}.
\newblock \showarticletitle{Access overlays: improving non-visual access to
  large touch screens for blind users}. In
  \bibinfo{booktitle}{\emph{Proceedings of the 24th annual {ACM} symposium on
  {User} interface software and technology - {UIST} '11}}.
  \bibinfo{publisher}{ACM Press}, \bibinfo{address}{Santa Barbara, California,
  USA}, \bibinfo{pages}{273}.
\newblock
\showISBNx{978-1-4503-0716-1}
\urldef\tempurl%
\url{https://doi.org/10.1145/2047196.2047232}
\showDOI{\tempurl}


\bibitem[\protect\citeauthoryear{Kaptein, Nass, and Markopoulos}{Kaptein
  et~al\mbox{.}}{2010}]%
        {kaptein_powerful_2010}
\bibfield{author}{\bibinfo{person}{Maurits~Clemens Kaptein},
  \bibinfo{person}{Clifford Nass}, {and} \bibinfo{person}{Panos Markopoulos}.}
  \bibinfo{year}{2010}\natexlab{}.
\newblock \showarticletitle{Powerful and consistent analysis of likert-type
  ratingscales}. In \bibinfo{booktitle}{\emph{Proceedings of the {SIGCHI}
  {Conference} on {Human} {Factors} in {Computing} {Systems}}}
  \emph{(\bibinfo{series}{{CHI} '10})}. \bibinfo{publisher}{Association for
  Computing Machinery}, \bibinfo{address}{Atlanta, Georgia, USA},
  \bibinfo{pages}{2391--2394}.
\newblock
\showISBNx{978-1-60558-929-9}
\urldef\tempurl%
\url{https://doi.org/10.1145/1753326.1753686}
\showDOI{\tempurl}


\bibitem[\protect\citeauthoryear{Kay}{Kay}{2020}]%
        {kay_contrast_2020}
\bibfield{author}{\bibinfo{person}{Matthew Kay}.}
  \bibinfo{year}{2020}\natexlab{}.
\newblock \bibinfo{title}{Contrast tests with {ART}}.
\newblock
\newblock
\urldef\tempurl%
\url{https://cran.r-project.org/web/packages/ARTool/vignettes/art-contrasts.html}
\showURL{%
\tempurl}


\bibitem[\protect\citeauthoryear{Kay, Nelson, and Hekler}{Kay
  et~al\mbox{.}}{2016}]%
        {kay_researcher-centered_2016}
\bibfield{author}{\bibinfo{person}{Matthew Kay}, \bibinfo{person}{Gregory~L.
  Nelson}, {and} \bibinfo{person}{Eric~B. Hekler}.}
  \bibinfo{year}{2016}\natexlab{}.
\newblock \showarticletitle{Researcher-{Centered} {Design} of {Statistics}:
  {Why} {Bayesian} {Statistics} {Better} {Fit} the {Culture} and {Incentives}
  of {HCI}}. In \bibinfo{booktitle}{\emph{Proceedings of the 2016 {CHI}
  {Conference} on {Human} {Factors} in {Computing} {Systems}}}
  \emph{(\bibinfo{series}{{CHI} '16})}. \bibinfo{publisher}{Association for
  Computing Machinery}, \bibinfo{address}{San Jose, California, USA},
  \bibinfo{pages}{4521--4532}.
\newblock
\showISBNx{978-1-4503-3362-7}
\urldef\tempurl%
\url{https://doi.org/10.1145/2858036.2858465}
\showDOI{\tempurl}


\bibitem[\protect\citeauthoryear{Krishnamoorthy}{Krishnamoorthy}{2016}]%
        {krishnamoorthy2016handbook}
\bibfield{author}{\bibinfo{person}{Kalimuthu Krishnamoorthy}.}
  \bibinfo{year}{2016}\natexlab{}.
\newblock \bibinfo{booktitle}{\emph{Handbook of statistical distributions with
  applications}}.
\newblock \bibinfo{publisher}{CRC Press}.
\newblock


\bibitem[\protect\citeauthoryear{Li}{Li}{[n.d.]}]%
        {li_robustness_nodate}
\bibfield{author}{\bibinfo{person}{Dong Li}.}
  \bibinfo{year}{[n.d.]}\natexlab{}.
\newblock \showarticletitle{Robustness {And} {Power} {Of} {The} {Student} {T},
  {Welch}-{Aspin}, {Yuen}, {Tukey} {Quick}, {And} {Haga} {Tests}}.
\newblock  (\bibinfo{year}{[n.\,d.]}), \bibinfo{pages}{1753}.
\newblock


\bibitem[\protect\citeauthoryear{Mann and Whitney}{Mann and Whitney}{1947}]%
        {mann_test_1947}
\bibfield{author}{\bibinfo{person}{H.~B. Mann} {and} \bibinfo{person}{D.~R.
  Whitney}.} \bibinfo{year}{1947}\natexlab{}.
\newblock \showarticletitle{On a {Test} of {Whether} one of {Two} {Random}
  {Variables} is {Stochastically} {Larger} than the {Other}}.
\newblock \bibinfo{journal}{\emph{The Annals of Mathematical Statistics}}
  \bibinfo{volume}{18}, \bibinfo{number}{1} (\bibinfo{year}{1947}),
  \bibinfo{pages}{50--60}.
\newblock
\showISSN{0003-4851}
\urldef\tempurl%
\url{https://www.jstor.org/stable/2236101}
\showURL{%
\tempurl}
\newblock
\shownote{Publisher: Institute of Mathematical Statistics.}


\bibitem[\protect\citeauthoryear{Mansouri}{Mansouri}{1998}]%
        {mansouri_multifactor_1998}
\bibfield{author}{\bibinfo{person}{H. Mansouri}.}
  \bibinfo{year}{1998}\natexlab{}.
\newblock \showarticletitle{Multifactor analysis of variance based on the
  aligned rank transform technique}.
\newblock \bibinfo{journal}{\emph{Computational Statistics \& Data Analysis}}
  \bibinfo{volume}{29}, \bibinfo{number}{2} (\bibinfo{date}{Dec.}
  \bibinfo{year}{1998}), \bibinfo{pages}{177--189}.
\newblock
\showISSN{01679473}
\urldef\tempurl%
\url{https://doi.org/10.1016/S0167-9473(98)00077-2}
\showDOI{\tempurl}


\bibitem[\protect\citeauthoryear{Mansouri}{Mansouri}{1999}]%
        {mansouri_aligned_1999}
\bibfield{author}{\bibinfo{person}{H. Mansouri}.}
  \bibinfo{year}{1999}\natexlab{}.
\newblock \showarticletitle{Aligned rank transform tests in linear models}.
\newblock \bibinfo{journal}{\emph{Journal of Statistical Planning and
  Inference}} \bibinfo{volume}{79}, \bibinfo{number}{1} (\bibinfo{date}{June}
  \bibinfo{year}{1999}), \bibinfo{pages}{141--155}.
\newblock
\showISSN{03783758}
\urldef\tempurl%
\url{https://doi.org/10.1016/S0378-3758(98)00229-8}
\showDOI{\tempurl}


\bibitem[\protect\citeauthoryear{Mansouri, Paige, and Surles}{Mansouri
  et~al\mbox{.}}{2004}]%
        {mansouri_aligned_2004}
\bibfield{author}{\bibinfo{person}{H. Mansouri}, \bibinfo{person}{R.~L. Paige},
  {and} \bibinfo{person}{J.~G. Surles}.} \bibinfo{year}{2004}\natexlab{}.
\newblock \showarticletitle{Aligned {Rank} {Transform} {Techniques} for
  {Analysis} of {Variance} and {Multiple} {Comparisons}}.
\newblock \bibinfo{journal}{\emph{Communications in Statistics - Theory and
  Methods}} \bibinfo{volume}{33}, \bibinfo{number}{9} (\bibinfo{date}{Dec.}
  \bibinfo{year}{2004}), \bibinfo{pages}{2217--2232}.
\newblock
\showISSN{0361-0926}
\urldef\tempurl%
\url{https://doi.org/10.1081/STA-200026599}
\showDOI{\tempurl}


\bibitem[\protect\citeauthoryear{Neave and Granger}{Neave and Granger}{1968}]%
        {neave_monte_1968}
\bibfield{author}{\bibinfo{person}{H.~R. Neave} {and} \bibinfo{person}{C.~W.~J.
  Granger}.} \bibinfo{year}{1968}\natexlab{}.
\newblock \showarticletitle{A {Monte} {Carlo} {Study} {Comparing} {Various}
  {Two}-{Sample} {Tests} for {Differences} in {Mean}}.
\newblock \bibinfo{journal}{\emph{Technometrics}} \bibinfo{volume}{10},
  \bibinfo{number}{3} (\bibinfo{year}{1968}), \bibinfo{pages}{509--522}.
\newblock
\showISSN{0040-1706}
\urldef\tempurl%
\url{https://doi.org/10.2307/1267105}
\showDOI{\tempurl}
\newblock
\shownote{Publisher: [Taylor \& Francis, Ltd., American Statistical
  Association, American Society for Quality].}


\bibitem[\protect\citeauthoryear{Olson}{Olson}{2013}]%
        {olson_efficacy_2013}
\bibfield{author}{\bibinfo{person}{Daryle~Alan Olson}.}
  \bibinfo{year}{2013}\natexlab{}.
\newblock \showarticletitle{The {Efficacy} {Of} {Select} {Nonparametric} {And}
  {Distribution}-{Free} {Research} {Methods}: {Examining} {The} {Case} {Of}
  {Concomitant} {Heteroscedasticity} {And} {Effect} {Of} {Treatment}}.
\newblock


\bibitem[\protect\citeauthoryear{Peterson}{Peterson}{2002}]%
        {peterson_six_2002}
\bibfield{author}{\bibinfo{person}{Kathleen Peterson}.}
  \bibinfo{year}{2002}\natexlab{}.
\newblock \showarticletitle{Six {Modifications} {Of} {The} {Aligned} {Rank}
  {Transform} {Test} {For} {Interaction}}.
\newblock \bibinfo{journal}{\emph{Journal of Modern Applied Statistical
  Methods}} \bibinfo{volume}{1}, \bibinfo{number}{1} (\bibinfo{date}{May}
  \bibinfo{year}{2002}), \bibinfo{pages}{100--109}.
\newblock
\showISSN{1538-9472}
\urldef\tempurl%
\url{https://doi.org/10.22237/jmasm/1020255240}
\showDOI{\tempurl}


\bibitem[\protect\citeauthoryear{Revilla-León, Jiang, Sadeghpour,
  Piedra-Cascón, Zandinejad, Özcan, and Krishnamurthy}{Revilla-León
  et~al\mbox{.}}{2019}]%
        {revilla-leon_intraoral_2019}
\bibfield{author}{\bibinfo{person}{Marta Revilla-León}, \bibinfo{person}{Peng
  Jiang}, \bibinfo{person}{Mehrad Sadeghpour}, \bibinfo{person}{Wenceslao
  Piedra-Cascón}, \bibinfo{person}{Amirali Zandinejad}, \bibinfo{person}{Mutlu
  Özcan}, {and} \bibinfo{person}{Vinayak~R. Krishnamurthy}.}
  \bibinfo{year}{2019}\natexlab{}.
\newblock \showarticletitle{Intraoral digital scans—{Part} 1: {Influence} of
  ambient scanning light conditions on the accuracy (trueness and precision) of
  different intraoral scanners}.
\newblock \bibinfo{journal}{\emph{The Journal of Prosthetic Dentistry}}
  (\bibinfo{date}{Dec.} \bibinfo{year}{2019}).
\newblock
\showISSN{0022-3913}
\urldef\tempurl%
\url{https://doi.org/10.1016/j.prosdent.2019.06.003}
\showDOI{\tempurl}


\bibitem[\protect\citeauthoryear{Richter}{Richter}{1999}]%
        {richter_nearly_1999}
\bibfield{author}{\bibinfo{person}{Scott~J. Richter}.}
  \bibinfo{year}{1999}\natexlab{}.
\newblock \showarticletitle{Nearly exact tests in factorial experiments using
  the aligned rank transform}.
\newblock \bibinfo{journal}{\emph{Journal of Applied Statistics}}
  \bibinfo{volume}{26}, \bibinfo{number}{2} (\bibinfo{date}{Feb.}
  \bibinfo{year}{1999}), \bibinfo{pages}{203--217}.
\newblock
\showISSN{0266-4763, 1360-0532}
\urldef\tempurl%
\url{https://doi.org/10.1080/02664769922548}
\showDOI{\tempurl}


\bibitem[\protect\citeauthoryear{Robertson and Kaptein}{Robertson and
  Kaptein}{2016}]%
        {robertson_modern_2016}
\bibfield{editor}{\bibinfo{person}{Judy Robertson} {and}
  \bibinfo{person}{Maurits Kaptein}} (Eds.). \bibinfo{year}{2016}\natexlab{}.
\newblock \bibinfo{booktitle}{\emph{Modern {Statistical} {Methods} for {HCI}}}.
\newblock \bibinfo{publisher}{Springer International Publishing},
  \bibinfo{address}{Cham}.
\newblock
\showISBNx{978-3-319-26631-2 978-3-319-26633-6}
\urldef\tempurl%
\url{https://doi.org/10.1007/978-3-319-26633-6}
\showDOI{\tempurl}


\bibitem[\protect\citeauthoryear{Roo and Hachet}{Roo and Hachet}{2017}]%
        {roo_one_2017}
\bibfield{author}{\bibinfo{person}{Joan~Sol Roo} {and} \bibinfo{person}{Martin
  Hachet}.} \bibinfo{year}{2017}\natexlab{}.
\newblock \showarticletitle{One {Reality}: {Augmenting} {How} the {Physical}
  {World} is {Experienced} by combining {Multiple} {Mixed} {Reality}
  {Modalities}}. In \bibinfo{booktitle}{\emph{Proceedings of the 30th {Annual}
  {ACM} {Symposium} on {User} {Interface} {Software} and {Technology}}}.
  \bibinfo{publisher}{ACM}, \bibinfo{address}{Québec City QC Canada},
  \bibinfo{pages}{787--795}.
\newblock
\showISBNx{978-1-4503-4981-9}
\urldef\tempurl%
\url{https://doi.org/10.1145/3126594.3126638}
\showDOI{\tempurl}


\bibitem[\protect\citeauthoryear{Salter and Fawcett}{Salter and
  Fawcett}{1993}]%
        {salter_1993_art}
\bibfield{author}{\bibinfo{person}{KC Salter} {and} \bibinfo{person}{RF
  Fawcett}.} \bibinfo{year}{1993}\natexlab{}.
\newblock \showarticletitle{The ART test of interaction: a robust and powerful
  rank test of interaction in factorial models}.
\newblock \bibinfo{journal}{\emph{Communications in Statistics-Simulation and
  Computation}} \bibinfo{volume}{22}, \bibinfo{number}{1}
  (\bibinfo{year}{1993}), \bibinfo{pages}{137--153}.
\newblock


\bibitem[\protect\citeauthoryear{Sawilowsky and Blair}{Sawilowsky and
  Blair}{1992}]%
        {sawilowsky_more_1992}
\bibfield{author}{\bibinfo{person}{Shlomo~S. Sawilowsky} {and}
  \bibinfo{person}{R.~Clifford Blair}.} \bibinfo{year}{1992}\natexlab{}.
\newblock \showarticletitle{A more realistic look at the robustness and {Type}
  {II} error properties of the t test to departures from population normality}.
\newblock \bibinfo{journal}{\emph{Psychological Bulletin}}
  \bibinfo{volume}{111}, \bibinfo{number}{2} (\bibinfo{date}{March}
  \bibinfo{year}{1992}), \bibinfo{pages}{352--360}.
\newblock
\showISSN{0033-2909}
\urldef\tempurl%
\url{https://doi.org/10.1037/0033-2909.111.2.352}
\showDOI{\tempurl}
\newblock
\shownote{Publisher: American Psychological Association.}


\bibitem[\protect\citeauthoryear{{Student}}{{Student}}{1908}]%
        {student_probable_1908}
\bibfield{author}{\bibinfo{person}{{Student}}.}
  \bibinfo{year}{1908}\natexlab{}.
\newblock \showarticletitle{The {Probable} {Error} of a {Mean}}.
\newblock \bibinfo{journal}{\emph{Biometrika}} \bibinfo{volume}{6},
  \bibinfo{number}{1} (\bibinfo{year}{1908}), \bibinfo{pages}{1--25}.
\newblock
\showISSN{0006-3444}
\urldef\tempurl%
\url{https://doi.org/10.2307/2331554}
\showDOI{\tempurl}
\newblock
\shownote{Publisher: [Oxford University Press, Biometrika Trust].}


\bibitem[\protect\citeauthoryear{Tukey}{Tukey}{1949}]%
        {tukey1949comparing}
\bibfield{author}{\bibinfo{person}{John~W Tukey}.}
  \bibinfo{year}{1949}\natexlab{}.
\newblock \showarticletitle{Comparing individual means in the analysis of
  variance}.
\newblock \bibinfo{journal}{\emph{Biometrics}} (\bibinfo{year}{1949}),
  \bibinfo{pages}{99--114}.
\newblock


\bibitem[\protect\citeauthoryear{Ware}{Ware}{1985}]%
        {ware_linear_1985}
\bibfield{author}{\bibinfo{person}{James~H. Ware}.}
  \bibinfo{year}{1985}\natexlab{}.
\newblock \showarticletitle{Linear {Models} for the {Analysis} of
  {Longitudinal} {Studies}}.
\newblock \bibinfo{journal}{\emph{The American Statistician}}
  \bibinfo{volume}{39}, \bibinfo{number}{2} (\bibinfo{year}{1985}),
  \bibinfo{pages}{95--101}.
\newblock
\showISSN{0003-1305}
\urldef\tempurl%
\url{https://doi.org/10.2307/2682803}
\showDOI{\tempurl}
\newblock
\shownote{Publisher: [American Statistical Association, Taylor \& Francis,
  Ltd.].}


\bibitem[\protect\citeauthoryear{Weisstein}{Weisstein}{2004}]%
        {weisstein2004bonferroni}
\bibfield{author}{\bibinfo{person}{Eric~W Weisstein}.}
  \bibinfo{year}{2004}\natexlab{}.
\newblock \showarticletitle{Bonferroni correction}.
\newblock \bibinfo{journal}{\emph{https://mathworld. wolfram. com/}}
  (\bibinfo{year}{2004}).
\newblock


\bibitem[\protect\citeauthoryear{Wilcoxon}{Wilcoxon}{1945}]%
        {wilcoxon_individual_1945}
\bibfield{author}{\bibinfo{person}{Frank Wilcoxon}.}
  \bibinfo{year}{1945}\natexlab{}.
\newblock \showarticletitle{Individual {Comparisons} by {Ranking} {Methods}}.
\newblock \bibinfo{journal}{\emph{Biometrics Bulletin}} \bibinfo{volume}{1},
  \bibinfo{number}{6} (\bibinfo{year}{1945}), \bibinfo{pages}{80--83}.
\newblock
\showISSN{0099-4987}
\urldef\tempurl%
\url{https://doi.org/10.2307/3001968}
\showDOI{\tempurl}
\newblock
\shownote{Publisher: [International Biometric Society, Wiley].}


\bibitem[\protect\citeauthoryear{Wilkinson and {The Task Force on Statistical
  Inference}}{Wilkinson and {The Task Force on Statistical Inference}}{1999}]%
        {wilkinson1999statistical}
\bibfield{author}{\bibinfo{person}{Leland Wilkinson} {and}
  \bibinfo{person}{{The Task Force on Statistical Inference}}.}
  \bibinfo{year}{1999}\natexlab{}.
\newblock \showarticletitle{Statistical methods in psychology journals:
  Guidelines and explanations}.
\newblock \bibinfo{journal}{\emph{American psychologist}} \bibinfo{volume}{54},
  \bibinfo{number}{8} (\bibinfo{year}{1999}), \bibinfo{pages}{594}.
\newblock


\bibitem[\protect\citeauthoryear{Wobbrock, Findlater, Gergle, and
  Higgins}{Wobbrock et~al\mbox{.}}{2011}]%
        {wobbrock_aligned_2011}
\bibfield{author}{\bibinfo{person}{Jacob~O. Wobbrock}, \bibinfo{person}{Leah
  Findlater}, \bibinfo{person}{Darren Gergle}, {and} \bibinfo{person}{James~J.
  Higgins}.} \bibinfo{year}{2011}\natexlab{}.
\newblock \showarticletitle{The {Aligned} {Rank} {Transform} for
  {Nonparametric} {Factorial} {Analyses} {Using} {Only} {Anova} {Procedures}}.
  In \bibinfo{booktitle}{\emph{Proceedings of the {SIGCHI} {Conference} on
  {Human} {Factors} in {Computing} {Systems}}} \emph{(\bibinfo{series}{{CHI}
  '11})}. \bibinfo{publisher}{ACM}, \bibinfo{address}{New York, NY, USA},
  \bibinfo{pages}{143--146}.
\newblock
\showISBNx{978-1-4503-0228-9}
\urldef\tempurl%
\url{https://doi.org/10.1145/1978942.1978963}
\showDOI{\tempurl}


\end{thebibliography}
